\documentclass[12pt, draftclsnofoot, onecolumn]{IEEEtran}
\usepackage{amsmath,amsthm,amsfonts,amscd,amssymb,graphics,dsfont}
\usepackage[mathscr]{eucal}
\usepackage{graphics,graphicx,multicol}
\usepackage{epsfig}
\usepackage{epstopdf}
\usepackage{subfigure}
\usepackage{stfloats}
\usepackage{latexsym}
\usepackage{amsfonts}
\usepackage{cite}
\usepackage{algorithm,algorithmic}
\usepackage{color}
\usepackage{xcolor}
\usepackage{multirow}

\newtheorem{prop}{Proposition}

\newtheorem{rem}{Remark}

\begin{document}
\bstctlcite{IEEEexample:BSTcontrol}
\title{Learning Optimal Fronthauling and Decentralized Edge Computation in\\Fog Radio Access Networks}
\author{Hoon Lee,~\IEEEmembership{Member,~IEEE}, Junbeom Kim,~\IEEEmembership{Student Member,~IEEE}, and Seok-Hwan Park,~\IEEEmembership{Member,~IEEE}\vspace{-15mm}
\thanks{H. Lee is with the Department of Smart Robot Convergence and Application Engineering and the Department of Information and Communications Engineering, Pukyong National University, Busan 48513, South Korea (e-mail: hlee@pknu.ac.kr).

J. Kim and S.-H. Park are with the Division of Electronic Engineering, Jeonbuk National University, Jeonju 54896, South Korea (e-mail: \{junbeom, seokhwan\}@jbnu.ac.kr).

© 2021 IEEE.  Personal use of this material is permitted.  Permission from IEEE must be obtained for all other uses, in any current or future media, including reprinting/republishing this material for advertising or promotional purposes, creating new collective works, for resale or redistribution to servers or lists, or reuse of any copyrighted component of this work in other works.}}
\maketitle
\begin{abstract}
Fog radio access networks (F-RANs), which consist of a cloud and multiple edge nodes (ENs) connected via fronthaul links, have been regarded as promising network architectures. The F-RAN entails a joint optimization of cloud and edge computing as well as fronthaul interactions, which is challenging for traditional optimization techniques. This paper proposes a Cloud-Enabled Cooperation-Inspired Learning (CECIL) framework, a structural deep learning mechanism for handling a generic F-RAN optimization problem. The proposed solution mimics cloud-aided cooperative optimization policies by including centralized computing at the cloud, distributed decision at the ENs, and their uplink-downlink fronthaul interactions. A group of deep neural networks (DNNs) are employed for characterizing computations of the cloud and ENs. The forwardpass of the DNNs is carefully designed such that the impacts of the practical fronthaul links, such as channel noise and signling overheads, can be included in a training step. As a result, operations of the cloud and ENs can be jointly trained in an end-to-end manner, whereas their real-time inferences are carried out in a decentralized manner by means of the fronthaul coordination. To facilitate fronthaul cooperation among multiple ENs, the optimal fronthaul multiple access schemes are designed. Training algorithms robust to practical fronthaul impairments are also presented. Numerical results validate the effectiveness of the proposed approaches.
\end{abstract}
\vspace{-5mm}
\begin{IEEEkeywords}
Deep learning, fog radio access networks, fronthaul interaction.
\end{IEEEkeywords}

\section{Introduction}
Centralized coordination of distributed edge nodes (ENs) has been brought great success in wireless communication networks \cite{SHPark:14,RTandon:16}. Such an architecture is realized with a cloud unit that schedules communication and computation of ENs by leveraging fronthaul interfaces. A particular example is a cloud radio access network (C-RAN) \cite{SPark:13,MTao:16,WLee:16,SHPark:18,JKim:19} where a cloud centrally performs the baseband signal processing, while radio-frequency (RF) functionalities are carried out by ENs, e.g., remote radio heads. 
The performance can be further enhanced by fog radio access networks (F-RANs) \cite{SHPark:16,JLiu:17,YXiao:18} where ENs are equipped with individual computing units.
Measurements of RF propagation environments, e.g., channel state information (CSI), are available only at the ENs due to the absence of the RF circuitry at the cloud. To perform centralized computations with distributed data, the cloud collects the local measurements through uplink fronthaul links.
The computation results of the cloud, which contain the information regarding the networking policies of the ENs, e.g., beamforming vectors, are forwarded via downlink fronthaul links. In the F-RAN systems, the data received from the cloud can be further processed at the ENs using local computing units. Hence, to optimize the F-RAN properly, we need to jointly design centralized cloud computing strategies, uplink-downlink fronthaul coordination, and distributed edge processing rules.

Recent studies \cite{SPark:13,MTao:16,WLee:16,SHPark:18,JKim:19,SHPark:16,JLiu:17,YXiao:18} have addressed various optimization tasks in the C-RAN and F-RAN systems. The works in \cite{SPark:13,MTao:16,WLee:16,SHPark:18,JKim:19,SHPark:16} have investigated a joint optimization of downlink fronthauling schemes at the cloud and multi-antenna signal processing at the ENs. Iterative algorithms are presented for tackling the nonconvexity of particular formulations. Assuming the capacity-constrained fronthauls, compression strategies of the cloud computing results
are determined along with the beamforming vectors at the ENs.
Although the downlink fronthaul interactions from the cloud to the ENs are adequately studied, they do not consider the imperfections occurred in the uplink fronthaul coordination such as the CSI update steps from the ENs to the cloud. Therefore, existing researches are suitable only for an ideal scenario where the global network state, e.g., the network CSIs, are perfectly known to the cloud.
Practical C-RAN systems should involve the joint optimization of downlink-uplink fronthauling protocols and centralized cloud computing strategies. It is, however, not trivial for traditional model-based optimization techniques \cite{SPark:13,MTao:16,WLee:16,SHPark:18,JKim:19,SHPark:16} since the fronthaul interactions typically invoke intractable features including random noises and fronthaul signaling designs.

For the F-RAN architecture, we need to additionally identify decentralized edge computation rules for individual ENs. Distributed optimization methods in the F-RAN have been studied in cache-enabled networks \cite{JLiu:17} and tactile Internet applications \cite{YXiao:18}. Message-passing algorithms are employed in \cite{JLiu:17} to determine a decentralized cache deployment policy. Each EN iteratively updates messages for the interactions with other ENs. These messages should be carefully designed for each network setup, thereby lacking the adaptivity as a general optimization framework.
The alternating direction method of multipliers (ADMM) approach can be exploited for the design of distributed and cooperative fog computing \cite{YXiao:18}. To facilitate iterative interactions among the ENs and the cloud, a proper reformulation technique is necessary to split a global optimization variable. These model-based decentralized algorithms cannot be straightforwardly applied to other types of optimization formulations. In addition, they do not take the practical fronthaul design issues into account such as quantization, noisy channels, and signaling overheads.

To overcome the drawbacks of traditional model-based algorithms, a {\em learning to optimize} paradigm has been intensively examined in various wireless networking scenarios \cite{HSun:18,WLee:18a,WLee:20,DLiu:20,JKim:20,PKerret:18,DGunduz:19,Kim2018,HLee:19b}. Deep neural networks (DNNs) are employed to replace unknown computation rules for solving network optimization problems.
Arbitrarily formulated objectives can be maximized in a data-driven manner without handcraft models, e.g., the convexity of functions and the prior information of the optimal solution. The DNNs are exploited to learn efficient power control mechanisms \cite{WLee:18a,WLee:20} and user association policy \cite{DLiu:20} in interfering wireless networks. Beamforming optimization problems for multi-antenna systems are addressed \cite{JKim:20}. These results reveal that deep learning (DL) approaches outperform existing suboptimal solutions with much reduced computational complexity. However, they are confined to centralized executions which are not suitable for the F-RAN systems.

Decentralized optimizations have been investigated via unsupervised DL \cite{DGunduz:19,Kim2018,HLee:19b} and reinforcement learning (RL) techniques \cite{Yasar:19}. In \cite{DGunduz:19,Kim2018,HLee:19b}, a distributed network setup is considered where direct interactions among ENs are allowed by leveraging backhaul interfaces. An interaction policy is autonomously optimized along with distributed computation rules. However, the setup in \cite{DGunduz:19,Kim2018,HLee:19b} is different from the F-RAN architecture where the ENs can only be controlled by the cloud, and thus they cannot optimize the role of the cloud in the F-RAN systems.
A cloud-aided distributed RL strategy is presented in \cite{Yasar:19}. To succeed the learning task, the RL framework requires a careful determination of state variables and rewards for individual ENs. The optimization of these hyperparameters typically incurs trial-and-error-based grid search procedures for each network setup. In addition, the backhaul imperfections and signaling overheads are not addressed in \cite{DGunduz:19,Kim2018,HLee:19b,Yasar:19}. Federated learning (FL) algorithms have been recently studied for handling distributed machine learning problems \cite{DGunduz:19,JPark:19}. The FL focuses on the training of a common DNN at the cloud with the aid of the ENs having individual training datasets. Thus, the FL would not be suitable for the design of decentralized optimization inferences in wireless networks where the ENs desire to identify their own networking solutions with partially observable statistics.

This paper proposes an unsupervised DL method for designing a generic optimization framework in the F-RAN systems. Distributed ENs observe their local states, e.g., the CSI for local wireless links, and desire to determine individual solutions, e.g., transmit power and beamforming vectors, for maximizing the network performance. Since the ENs are typically deployed in a wide cell coverage area, the locally observable information of a certain EN is not directly available to others. A network cloud connecting the ENs through imperfect fronthaul links schedules decentralized edge processing. To optimize the operations at the cloud and the ENs jointly, we propose a Cloud-Enabled Cooperation-Inspired Learning (CECIL) mechanism, which is a structural DL solution developed for the F-RAN systems. The proposed method consists of three consecutive steps: uplink fronthauling at ENs, centralized computation and downlink fronthauling at a cloud, and distributed decision at ENs.
A group of DNN units is employed for characterizing the operations of the cloud and ENs. A joint training algorithm of the DNNs is presented with arbitrary given fronthaul imperfections.

The uplink and downlink fronthaul interaction steps incur inter-EN inference signals. To handle this issue, we design multiple access fronthauling schemes that can be autonomously optimized by the DNNs. Two different protocols are investigated. First, following conventional distributed DL approaches \cite{DGunduz:19,Kim2018,HLee:19b,Yasar:19}, an orthogonal multiple access (OMA) is presented which assigns distinct fronthaul resources for each EN. Second, we propose a non-orthogonal multiple access (NOMA) fronthauling strategy where all ENs share the identical fronthaul resources. The non-orthogonal interaction policies among ENs have not yet been investigated in existing DL studies \cite{DGunduz:19,Kim2018,HLee:19b,Yasar:19}, and thus its optimality would not be guaranteed in the design of the cooperative DNN inferencing steps. To this end, we rigorously prove the effectiveness of the OMA and NOMA schemes and analyze the amount of the fronthaul resources to achieve the optimality. The superiority of the NOMA method is verified in terms of the fronthaul signaling overheads. In addition, for the imperfect fronthaul link case, we present a robust learning policy that trains the DNNs in the presence of practical fronthaul impairments such as additive noise and finite-capacity constraints. Finally, numerical results verify the effectiveness of the proposed framework in various F-RAN applications. Our main contributions are summarized as follows.
\begin{itemize}
\item We propose the CECIL framework, a model-driven DL-based optimization mechanism for the F-RAN structure, which jointly determines the decentralized edge computations, centralized cloud calculations, and uplink-downlink fronthaul coordination strategies.
\item For managing inter-edge interference, fronthaul multiple access schemes are designed which bridge computations of DNNs at the cloud and ENs. The optimality of the proposed fronthauling strategies are verified rigorously.
\item To combat the fronthaul channel imperfections, robust training policies are presented which optimize DNNs in the presence of additive noise and fronthaul capacity constraints.
\item Intensive numerical results validating the optimality of the proposed method are provided in interfering networks. Efficient fronthaul resource allocation methods are identified from the numerical results.
\end{itemize}

The rest of the paper is organized as follows. Section~\ref{sec:sec2} describes a generic F-RAN system. The inference of the CECIL framework is explained in Section \ref{sec:sec3}, and its training process is presented in Section \ref{sec:sec4}. Optimal fronthaul interaction strategies are designed in Section \ref{sec:sec5}, and in Section \ref{sec:sec6}, robust training policies for imperfect fronthaul channels are studied. Section \ref{sec:sec7} assesses the performance of the proposed CECIL approach from numerical simulations. Finally, concluding remarks are given in Section \ref{sec:sec8}.


\section{Network Model and Problem Formulation} \label{sec:sec2}

\begin{figure}
\centering
\includegraphics[width=.5\linewidth]{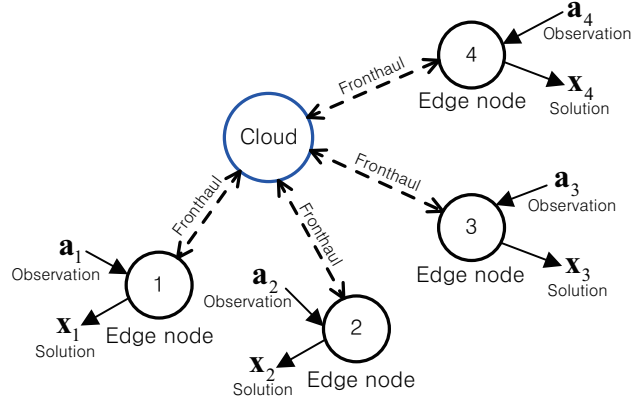}
\caption{F-RAN system with a cloud and $N=4$ ENs.}
\label{fig:fig1}
\vspace{-3mm}
\end{figure}

Fig. \ref{fig:fig1} illustrates an F-RAN architecture which exploits both cloud and edge computing processes for an efficient management of wireless networks. A cloud is regarded as a central unit that coordinates multiple, say $N$, ENs by means of fronthaul links. The ENs are equipped with RF modules to provide networking services. We maximize a generic nonconvex network utility function $f(\cdot)$ by optimizing network policies at the ENs. Without loss of generality, states of the F-RAN are represented by a vector $\mathbf{a}\in\mathbb{R}^{A}$ of length $A$. The global state $\mathbf{a}$ can be any measurement values such as a set of CSIs between the ENs and their intended mobile users. The ENs equipped with the RF processors are responsible for the estimation of the global state vector. The ENs are, in general, distributed over coverage areas to support reliable communication services. Therefore, a locally observable state at each EN $i$ ($i=1,\cdots,N$) denoted by $\mathbf{a}_{i}\in\mathbb{R}^{A_{i}}$ becomes a subset of $\mathbf{a}$ and it is not known to other ENs and the cloud. Then, the global information vector $\mathbf{a}$ can be represented by $\mathbf{a}\triangleq\{\mathbf{a}_{i}:\forall i\}$ with size $A\triangleq\sum_{i=1}^{N}A_{i}$.

The fronthaul interface supports the cooperation among the cloud and ENs.
For notational simplicity, the cloud is denoted by the $0$-th node. Let $M_{i0}$ and $M_{0i}$ be the number of uplink and downlink fronthaul resource blocks (RBs) assigned for EN $i$, respectively. Without loss of the generality, one fronthaul RB is assumed to be occupied for conveying a real-valued scalar number. In practice, the RBs correspond to orthogonal time-frequency channels, e.g., a resource element in LTE systems which consists of one data symbol occupying 15 kHz bandwidth. Thus, $M_{i0}$ and $M_{0i}$ reflect the fronthaul signaling overheads and the fronthaul resource constraints. The capacity constraints on the fronthaul RBs are addressed in Section \ref{sec:sec6B}. The total available RBs for the F-RAN is limited by $M$ as $M_{i0}\leq M$ and $M_{0i}\leq M$. The number of the RBs can be optimized in advance by the network operator and is assumed to be fixed. The RB allocation schemes are discussed in Section \ref{sec:sec7A}. Both time division duplexing (TDD) and frequency division duplexing (FDD) protocols can be exploited for implementing the fronthaul coordinations. For the TDD systems, we have $M_{i0}=M_{0i}$ since the quality of the uplink and downlink fronthaul channels is the same due to the channel reciprocity. Also, a more general case of $M_{i0}\neq M_{0i}$ represents the FDD systems where the uplink and downlink fronthaul transmissions experience different radio propagation environments. As a result, the DL method presented in the preceding sections can be applied to arbitrary duplexing systems including both the TDD and FDD.


A decision of EN $i$ is characterized by a solution vector $\mathbf{x}_{i}\in\mathbb{R}^{X_{i}}$ of length $X_{i}$ which includes resource management policy and beamforming vector of EN $i$. The performance of the F-RAN is generally affected by both the global state $\mathbf{a}$ and a set of solutions $\mathbf{x}\triangleq\{\mathbf{x}_{i}:\forall i\}$. Thus, the utility function can be written by $f(\mathbf{a},\mathbf{x})$. We focus on a maximization task of the utility averaged over the global state vector $\mathbf{a}$ expressed by
\begin{align}
(\text{P}1): &\max_{\mathbf{x}}\mathbb{E}_{\mathbf{a}}[f(\mathbf{a},\mathbf{x})]\nonumber\\
&\text{subject to }\mathbf{x}_{i}\in\mathcal{D}_{i},\forall i\nonumber
\end{align}
where $\mathbb{E}_{U}[\cdot]$ is the expectation operation over a random variable $U$ and $\mathcal{D}_{i}$ stands for a solution set of EN~$i$. To tackle (P1) in the F-RAN system, along with the solution vector $\mathbf{x}$, we need to identify the fronthaul interaction policy subject to the fronthaul RB constraints $M_{i0}$ and $M_{0i}$, $\forall i$. The effect of the fronthaul noise and inter-EN interference can be included in (P1). These are distinct features of our formulation (P1) compared to existing studies on decentralized  multi-agent architectures \cite{Yasar:19} which do not consider the resource constraints on the coordination links.

In this paper, we develop an efficient solution for a generic formulation (P1) whose computational inferences can be realized in the F-RAN systems.
Major challenges for (P1) arise from the distinctly available observations and imperfect fronthaul interfaces. We need to perfectly know $\mathbf{a}$ to solve (P1).
One possible approach is to let the ENs upload their local measurements $\mathbf{a}_{i}$ to the cloud through the uplink fronthaul links. Then, the cloud can calculate the network solution $\mathbf{x}$. The local decision variables $\mathbf{x}_{i}$ are transferred to individual ENs by leveraging the downlink fronthaul links. Such a strategy is only applicable to an ideal scenario where the fronthaul links are perfect and have sufficient RBs, i.e., $M_{i0}\geq A_{i}$ and $M_{0i}\geq X_{i}$ for exchanging $\mathbf{a}_{i}\in\mathbb{R}^{A_{i}}$ and $\mathbf{x}_{i}\in\mathbb{R}^{X_{i}}$, respectively. Conventional approaches \cite{SPark:13,MTao:16,WLee:16,SHPark:18,JKim:19,SHPark:16} only focus on the downlink or the uplink compression, but not their joint design. 
The existing FL algorithms \cite{DGunduz:19,JPark:19}, where the ENs cooperatively find a common solution at the cloud via model-based iteration rules, are not suitable for addressing the F-RAN problem (P1) since it requires to optimize fronthaul interaction policies. To efficiently solve (P1) in the F-RAN system, we need to study a joint design of fronthaul communication policies and individual decisions at the ENs that can be applied to arbitrary utility functions.

\section{Cooperative Learning Mechanism}\label{sec:sec3}
\begin{figure}
\centering
    \subfigure[Uplink message generation]{
        \includegraphics[width=.33\linewidth]{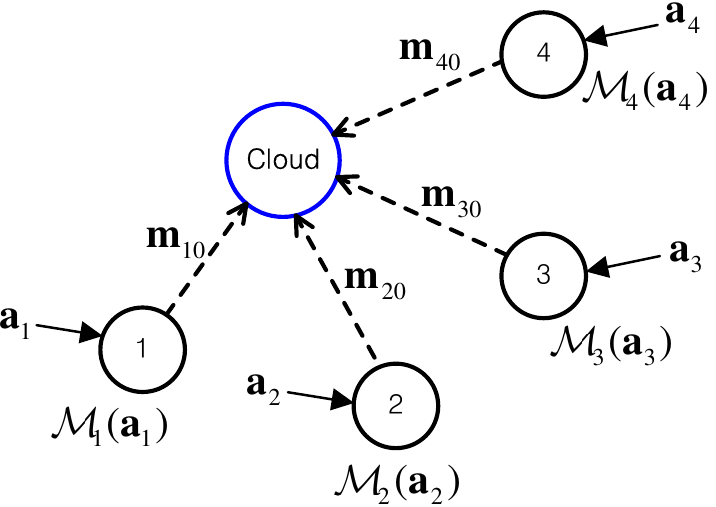}\label{fig:fig2a}
    }
    \subfigure[Downlink message generation]{
        \includegraphics[width=.24\linewidth]{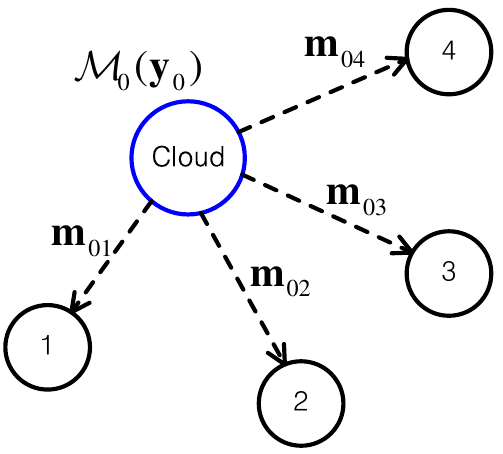}\label{fig:fig2b}
    }
    \subfigure[Distributed decision]{
        \includegraphics[width=.35\linewidth]{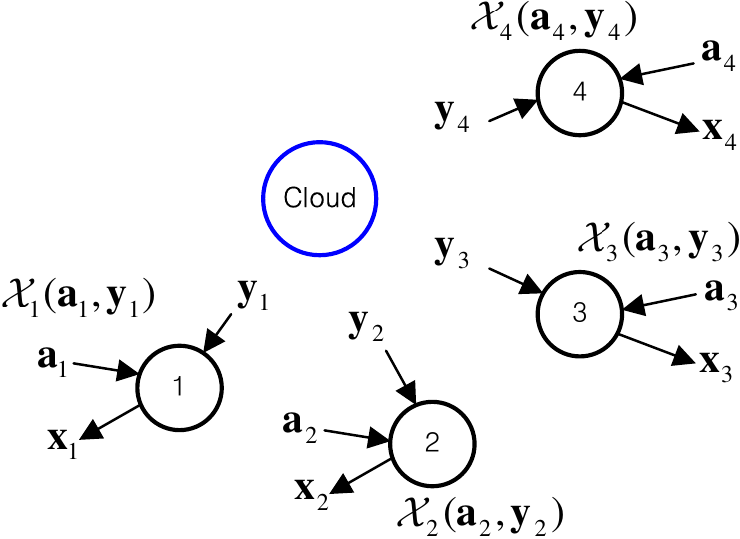}\label{fig:fig2c}
    }
    \caption{Proposed cooperative inference.}
    \label{fig:fig2}
\end{figure}

This section presents a CECIL inference which designs cooperative optimization mechanisms for the F-RAN system.
The CECIL is exploited as forward pass computations of a DNN-based optimization framework in Section \ref{sec:sec4}.
We characterize interactions among the cloud and ENs by leveraging abstracted computational inferences to be replaced by DNNs. Fig. \ref{fig:fig2} describes the proposed cooperative inference structure which consists of three sequential steps: uplink message generation at the ENs, downlink message generation at the cloud, and the decentralized decision at the ENs. In what follows, we describe the details of each step.

\subsection{Uplink message generation at ENs}
As shown in Fig. \ref{fig:fig2a}, EN $i$ first sends the information regarding its local observation $\mathbf{a}_{i}$ to the cloud using the uplink fronthaul link assigned with $M_{i0}$ RBs. A straightforward transmission of the $A_{i}$-dimensional raw data $\mathbf{a}_{i}$ would not be possible when we have insufficient fronthaul resources as $M_{i0}\leq A_{i}$. Thus, EN $i$ needs to identify a low-dimensional representation of $\mathbf{a}_{i}$ without no direct interactions with other ENs. This can be viewed as decentralized edge compression steps. The resulting representation $\mathbf{m}_{i0}\in\mathbb{R}^{M_{i0}}$ of length $M_{i0}$ is referred to as an {\em uplink message} that carries the local knowledge of EN $i$ to the cloud via $M_{i0}$ fronthaul RBs. Let $\mathcal{M}_{i}(\cdot)$ be a computational inference performing the uplink message generation of EN $i$,~i.e.,
\begin{align}
\mathbf{m}_{i0}=\mathcal{M}_{i}(\mathbf{a}_{i}).\label{eq:mi0}
\end{align}
In \eqref{eq:mi0}, only the local observation $\mathbf{a}_{i}$ is accepted as an input for characterizing fully decentralized processing. As discussed in Section \ref{sec:sec4}, the inference $\mathcal{M}_{i}(\cdot)$ is modeled by a DNN to be optimized for maximizing the utility.

\subsection{Downlink message generation at cloud}
Practical fronthaul links are interrupted with channel impairments such as the noise, and thus the cloud would get the noisy observation of the uplink messages. To capture this, we introduce a channel transfer function $h_{i0}(\cdot)$ for the uplink fronthaul link from EN $i$ to the cloud which can include any channel imperfection encountered in the uplink communication. Then, the received signal at the cloud $\mathbf{y}_{0}$ depends on all the noisy uplink messages $h_{i0}(\mathbf{m}_{i0})$, $\forall i$. It is written by
\begin{align}
\mathbf{y}_{0}=u\big(\{h_{i0}(\mathbf{m}_{i0}):\forall i\}\big)=u\big(\{h_{i0}(\mathcal{M}_{i}(\mathbf{a}_{i})):\forall i\}\big),\label{eq:y0}
\end{align}
where the function $u(\cdot)$ defined over the set of the noisy messages $\{h_{i0}(\mathbf{m}_{i0}):\forall i\}$ describes an uplink transmission strategy of the ENs. The choice of $u(\cdot)$ relies on the fronthaul resource sharing policy. For instance, if each EN occupies distinct fronthaul RBs, $u(\cdot)$ is simply given by the concatenation operation. On the other hand, $u(\cdot)$ becomes the summation when all ENs share the entire uplink fronthaul RBs. The dimension of $\mathbf{y}_{0}$ depends on the number of the uplink fronthaul RBs $M_{i0}$ and the uplink signaling strategy $u(\cdot)$. These are specified in Section~\ref{sec:sec5}.

From \eqref{eq:y0}, we can observe that the received signal $\mathbf{y}_{0}$ conveys distorted information of the global observation $\mathbf{a}=\{\mathbf{a}_{i}:\forall i\}$ with the piecewise edge processing $\mathcal{M}_{i}(\cdot)$. A standard approach to process with $\mathbf{y}_{0}$ is to decompose the computations of the cloud into the following subsequent steps. The cloud first recovers the global state $\mathbf{a}$ from the received signal $\mathbf{y}_{0}$. Then, the solution to (P1) is determined by centralized cloud computing strategies. The resulting solution $\mathbf{x}_{i}\in\mathbb{R}^{X_{i}}$ is sent back to EN $i$ via the downlink fronthaul links with $M_{0i}$ RBs. To handle the practical case with $M_{0i}\leq X_{i}$, $\mathbf{x}_{i}$ is encoded into a {\em downlink message} $\mathbf{m}_{0i}\in\mathbb{R}^{M_{0i}}$ whose dimension $M_{0i}$ is fit to the number of the downlink fronthaul RBs $M_{0i}$ assigned to EN $i$. As illustrated in Fig. \ref{fig:fig2b}, we integrate such cascaded procedures into a single computation inference $\mathcal{M}_{0}(\cdot)$ that creates a set of the downlink messages $\{\mathbf{m}_{0i}:\forall i\}$ from $\mathbf{y}_{0}$. This can be written~by
\begin{align}
\{\mathbf{m}_{0i}:\forall i\}=\mathcal{M}_{0}(\mathbf{y}_{0}).\label{eq:m0i}
\end{align}
It is inferred from \eqref{eq:m0i} that the downlink message $\mathbf{m}_{0i}$ encapsulates the local observations of other nodes $\mathbf{a}_{j}$, $\forall j\neq i$, as well as an intermediate decision taken at the cloud. The inference $\mathcal{M}_{0}(\cdot)$ is also modeled by a DNN whose parameters are determined to maximize the utility function.

\begin{rem}
The inference in \eqref{eq:m0i} can be viewed as a two-way relaying strategy \cite{RZhang:09b,KJLee:10} where the cloud relays the signals received from the ENs after an appropriate signal processing $\mathcal{M}_{0}(\cdot)$. Classical relaying protocols are dependent on man-made signaling strategies, e.g., amplify-and-forward and decode-and-forward \cite{NLaneman:04}, which might not be the optimum cooperation policy. The proposed DL approach can identify the optimal relaying protocol, provided that a relaying inference $\mathcal{M}_{0}(\cdot)$ is approximated by a properly constructed DNN.
\end{rem}

\subsection{Distributed decision at ENs}
The distributed decision process shown in Fig. \ref{fig:fig2c} is described. The cloud broadcasts the downlink messages to EN $i$ with a pre-designed downlink signaling strategy denoted by $d_{i}(\cdot)$. Similar to the uplink signaling strategy $u(\cdot)$ in \eqref{eq:y0}, $d_{i}(\cdot)$ is defined over the set of the downlink messages $\{\mathbf{m}_{0i}:\forall i\}$ and becomes a design factor to be specified in Section \ref{sec:sec5}. The downlink signal intended to EN $i$, which is denoted by $\mathbf{d}_{i}$, can be written by
\begin{align}
\mathbf{d}_{i}=d_{i}(\{\mathbf{m}_{0j}:\forall j\}).\label{eq:d}
\end{align}
Defining $h_{0i}(\cdot)$ as the downlink fronthaul transfer function from the cloud to EN $i$, the received signal $\mathbf{y}_{i}$ at EN $i$ is given by
\begin{align}
\mathbf{y}_{i}=h_{0i}(\mathbf{d}_{i})=h_{0i}\big(d_{i}(\{\mathbf{m}_{0j}:\forall j\})\big).\label{eq:yi}
\end{align}
The dimensions of $\mathbf{d}_{i}$ and $\mathbf{y}_{i}$ rely on the message broadcasting strategy $d_{i}(\cdot)$ to be designed in Section \ref{sec:sec5}. Combining \eqref{eq:mi0}, \eqref{eq:m0i}, and \eqref{eq:yi}, we can see that the received message $\mathbf{y}_{i}$ of EN $i$ contains the local statistics of all the ENs. This implies that all the sufficient, but possibly corrupted, information for solving (P1) is now available at each EN. Thereby, the solution $\mathbf{x}_{i}$ of EN $i$ can be attained individually by means of a node-centric decision inference $\mathcal{X}_{i}(\cdot)$. The proposed solution computation rule at EN $i$ is expressed as
\begin{align}
\mathbf{x}_{i}=\mathcal{X}_{i}(\mathbf{a}_{i},\mathbf{y}_{i}).\label{eq:xi}
\end{align}
We use the local observation $\mathbf{a}_{i}$ as the side information to refine the received signal $\mathbf{y}_{i}$ dedicated to EN $i$. This additional input forms a residual shortcut which leads to an efficient training strategy of very deep networks \cite{KHe:16}.

\begin{algorithm}
\caption{Proposed CECIL inference for F-RAN}
\begin{algorithmic}\label{alg:alg1}
    \STATE {\em 1. Uplink message generation:}\\
    ~~EN $i$, $\forall i$, creates an uplink message $\mathbf{m}_{i0}$ from \eqref{eq:mi0} and sends it to the cloud through the uplink fronthaul links~\eqref{eq:y0}.
    \STATE {\em 2. Downlink message generation:}\\
    ~~The cloud broadcasts downlink messages $\mathbf{m}_{0i}$ generated from \eqref{eq:m0i} using the downlink fronthaul links~\eqref{eq:yi}.
    \STATE {\em 3. Distributed decision:}\\
    ~~EN $i$, $\forall i$, computes an individual solution $\mathbf{x}_{i}$ from~\eqref{eq:xi}.
\end{algorithmic}
\end{algorithm}

Algorithm~\ref{alg:alg1} summarizes the inference of the CECIL framework. The uplink messages generated at the ENs are first transmitted to the cloud. Receiving the noisy signal $\mathbf{y}_{0}$, the centralized cloud computing yields the downlink messages to be broadcasted to the ENs. The decision $\mathbf{x}_{i}$ is then taken at each EN $i$ individually. The proposed inference relies only on locally observable information, i.e., local measurement $\mathbf{a}_{i}$ and the received messages, but not on instantaneous states of other network entities. As a result, Algorithm \ref{alg:alg1} can be implemented in a distributed manner with optimized $\mathcal{M}_{i}(\cdot)$ and $\mathcal{X}_{i}(\cdot)$.

\section{Deep Learning Formulation}\label{sec:sec4}
Based on the formulations in \eqref{eq:mi0}, \eqref{eq:m0i}, and \eqref{eq:xi}, the original problem (P1) can be transformed~as
\begin{align}
(\text{P}2): &\max_{\{\mathcal{M}_{i}(\cdot),\mathcal{X}_{i}(\cdot):\forall i\}}
    \mathbb{E}_{\mathbf{a}}[f(\mathbf{a},\{\mathcal{X}_{i}(\mathbf{a}_{i},\mathbf{y}_{i}):\forall i\})]\nonumber\\
    &~~~~\text{subject to }\mathcal{X}_{i}(\mathbf{a}_{i},\mathbf{y}_{i})\in\mathcal{D}_{i},\forall i,\ \forall\mathbf{a}.\nonumber
\end{align}
The targets of the optimization are given by unstructured functions $\mathcal{M}_{i}(\cdot)$ in \eqref{eq:mi0}, $\mathcal{M}_{0}$ in \eqref{eq:m0i}, and $\mathcal{X}_{i}(\cdot)$ in \eqref{eq:xi}, which cannot be tackled by traditional optimization techniques requiring analytical formulas. To this end, we employ the learning to optimize approach \cite{HSun:18,WLee:18a,WLee:20,DLiu:20,JKim:20,PKerret:18,DGunduz:19,Kim2018,HLee:19b} which employs DNNs for replacing unknown mappings $\mathcal{M}_{i}(\cdot)$ and $\mathcal{X}_{i}(\cdot)$.
Let $\mathcal{F}_{Q}(\cdot;\theta)$ be a $Q$-layer fully-connected DNN with a trainable parameter $\mathbf{\theta}$. For an input vector $\mathbf{u}\in\mathbb{R}^{U_{1}}$ of length $U_{1}$, the output of $\mathcal{F}_{Q}(\mathbf{u};\theta)$ is written~as
\begin{align}
\mathcal{F}_{Q}(\mathbf{u};\mathbf{\theta})=\sigma_{Q}(\mathbf{W}_{Q}\!\times\!\cdots\!\times\!\sigma_{1}(\mathbf{W}_{1}\mathbf{u}+\mathbf{b}_{1})\!+\!\cdots\!+\!\mathbf{b}_{Q}),\label{eq:fnn}
\end{align}
where $\sigma_{q}(\cdot)$ is an activation at layer $q~(q=1,\cdots,Q)$ and $\mathbf{\theta}$ accounts for the collection of weight matrices $\mathbf{W}_{q}\in\mathbb{R}^{U_{q+1}\times U_{q}}$ and bias vectors $\mathbf{b}_{q}\in\mathbb{R}^{U_{q}}$ for all layers.

We replace the mappings in \eqref{eq:mi0}, \eqref{eq:m0i}, and \eqref{eq:xi} with DNNs~as
\begin{align}
\mathbf{m}_{i0}&=\mathcal{M}_{i}(\mathbf{a}_{i})=\mathcal{F}_{Q_{M_{i}}}(\mathbf{a}_{i};\theta_{M_{i}}),\label{eq:mi0_fnn}\\
\{\mathbf{m}_{0i}:\forall i\}&=\mathcal{M}_{0}(\mathbf{y}_{0})=\mathcal{F}_{Q_{M_{0}}}(\mathbf{y}_{0};\theta_{M_{0}}),\label{eq:m0i_fnn}\\
\mathbf{x}_{i}&=\mathcal{X}_{i}(\mathbf{a}_{i},\mathbf{y}_{i})=\mathcal{F}_{Q_{X_{i}}}(\mathbf{a}_{i}\oplus\mathbf{y}_{i};\theta_{X_{i}}),\label{eq:xi_fnn}
\end{align}
where $\mathbf{u}\oplus\mathbf{v}\triangleq[\mathbf{u}^{T},\mathbf{v}^{T}]^{T}$ stands for a concatenation operation of two vectors $\mathbf{u}$ and $\mathbf{v}$. The output dimension of $\mathcal{F}_{Q_{M_{i}}}(\cdot;\theta_{M_{i}})$, $\mathcal{F}_{Q_{M_{0}}}(\cdot;\theta_{M_{0}})$, and $\mathcal{F}_{Q_{X_{i}}}(\cdot;\theta_{X_{i}})$ are respectively set to the lengths of the desired outputs. The optimality of this DNN approximation is guaranteed by the universal approximation theorem \cite{KHornik:89}. It states that for any continuous mapping $z(\mathbf{u})$ defined on a compact set $\mathbf{u}\in\mathcal{U}$, there exist a finite $Q$ and arbitrary small $\varepsilon>0$ such that
\begin{align}
\sup_{\mathbf{u}\in\mathcal{U}}\|z(\mathbf{u})-\mathcal{F}_{Q}(\mathbf{u};\mathbf{\theta})\|\leq\varepsilon \label{eq:univ1}
\end{align}
with $\varepsilon>0$ being an arbitrary small number. From \eqref{eq:univ1}, we can identify a DNN close to any continuous function in terms of the worst-case Euclidean distance. Note that \eqref{eq:univ1} also holds for the unknown optimal mappings $\mathcal{M}_{i}^{\star}(\cdot)$ and $\mathcal{X}_{i}^{\star}(\cdot)$. Therefore, the DNN approximations in \eqref{eq:mi0_fnn}-\eqref{eq:xi_fnn} can provide a tractable formulation of (P2) but without loss of the optimality.

\begin{figure}
\centering
\includegraphics[width=.8\linewidth]{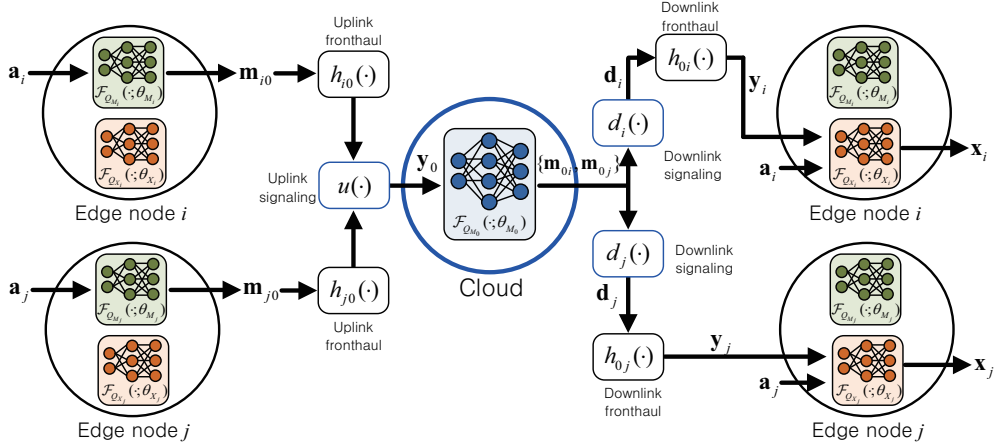}
\caption{End-to-end structure of proposed CECIL-based F-RAN system.}
\label{fig:fig3}
\end{figure}

\subsection{Training and Implementation}
Fig. \ref{fig:fig3} illustrates the CECIL-based F-RAN systems where the computations of the ENs and cloud are carried out by the DNNs in \eqref{eq:mi0_fnn}-\eqref{eq:xi_fnn}. The forward pass computations of the CECIL are provided in Algorithm \ref{alg:alg1}.
Plugging \eqref{eq:mi0_fnn}-\eqref{eq:xi_fnn} to (P2) results~in
\begin{align}
(\text{P}3): &\max_{\boldsymbol{\Theta}}
    \mathbb{E}_{\mathbf{a}}[f(\mathbf{a},\{\mathcal{F}_{Q_{X_{i}}}(\mathbf{a}_{i}\oplus\mathbf{y}_{i};\theta_{X_{i}}):\forall i\})]\nonumber\\
    &~\text{subject to }\mathcal{F}_{Q_{X_{i}}}(\mathbf{a}_{i}\oplus\mathbf{y}_{i};\theta_{X_{i}})\in\mathcal{D}_{i},\forall i,\ \forall\mathbf{a},\nonumber
\end{align}
where $\boldsymbol{\Theta}$ accounts for the set of learnable parameters of the DNNs in \eqref{eq:mi0_fnn}-\eqref{eq:xi_fnn} defined as
\begin{align}
\boldsymbol{\Theta}\triangleq\{\theta_{M_{i}}:\forall i=0,1,\cdots,N\}\bigcup\{\theta_{X_{i}}:\forall i=1,\cdots,N\}.
\end{align}
To remove the constraint of (P2), the output activation of $\mathcal{F}_{Q_{X_{i}}}(\cdot;\theta_{X_{i}})$ can be designed as the projection operator $\arg\min_{\mathbf{v}\in\mathcal{D}_{i}}\|\mathbf{u}-\mathbf{v}\|$ for a layer input $\mathbf{u}$. For the convex feasibility set $\mathcal{D}_{i}$, this projection activation is given by a convex quadratic program (QP) whose gradient-based training rules can be obtained with the backpropagation algorithm~\cite{BAmos:17}. The nonconvex projection problem can be tackled by the successive convex approximation mechanism \cite{YSun:17} by solving a series of approximated convex QPs. The gradients of such an iterative procedure can be obtained by integrating the gradients of approximated convex QPs. As a consequent, (P2) is readily solved by the gradient descent method and its variants for stochastic optimizations, e.g., the Adam algorithm \cite{Kingma:15}.
We adopt the mini-batch stochastic gradient descent (SGD) method \cite{IGoodfellow:16} where the expectations over the distribution of $\mathbf{a}$ are estimated as the sample mean evaluated on the mini-batch sets $\mathcal{A}\triangleq\{\mathbf{a}\}$. The SGD update at the $t$-th training epoch is given by
\begin{align}
\boldsymbol{\Theta}^{(t)}=\boldsymbol{\Theta}^{(t-1)}+\alpha\frac{1}{|\mathcal{A}|}\sum_{\mathbf{a}\in\mathcal{A}}\nabla_{\boldsymbol{\Theta}}f(\mathbf{a},\{\mathcal{F}_{Q_{X_{i}}}(\mathbf{a}_{i}\oplus\mathbf{y}_{i};\theta_{X_{i}}^{(t-1)}):\forall i\}),\label{eq:sgd}
\end{align}
where $q^{(t)}$ indicates a variable $q$ attained at the $t$-th epoch, $\alpha>0$ is a learning rate, and $\nabla_{q}$ denotes the gradient operator with respect to $q$.
The sample gradient $\nabla_{\boldsymbol{\Theta}}f(\mathbf{a},\{\mathcal{F}_{Q_{X_{i}}}(\mathbf{a}_{i}\oplus\mathbf{y}_{i};\theta_{X_{i}}:\forall i\})$ can be numerically calculated by the backpropagation algorithm \cite{IGoodfellow:16}, provided that the gradients of the channels $h_{i0}(\cdot)$ and $h_{0i}(\cdot)$ as well as the transmission strategies $u(\cdot)$ and $d_{i}(\cdot)$ are available. 

The full knowledge of the global observation $\mathbf{a}$ is required for computing the gradient of the utility function. This can be achieved by the centralized training procedure in an offline domain before real-time optimization inferences \cite{PKerret:18,DGunduz:19,Kim2018,HLee:19b}. To this end, we can collect training samples, i.e., a set of the local observation vectors $\mathbf{a}_{i}$, from the ENs in advance.
No labels such as the information regarding the optimal solution to (P1) are needed in the training. Thus, the proposed training strategy \eqref{eq:sgd} is performed in a fully unsupervised manner. Once the parameter set $\boldsymbol{\Theta}$ is determined, they are readily implemented at the cloud and ENs. As discussed, the forward pass in Algorithm \ref{alg:alg1} can be carried out only with locally measurable statistics, thereby leading to the distributed realization of the online computations \eqref{eq:mi0_fnn}-\eqref{eq:xi_fnn}. Compared to existing decentralized F-RAN optimization algorithms \cite{JLiu:17,YXiao:18} that require iterative procedures, the proposed CECIL does not need any repetitions in the real-time inference step. Hence, the proposed approach can save both the fronthaul signaling and computation overheads.

\section{Message Multiple Access Design}\label{sec:sec5}
The uplink and downlink interaction steps involve the transmission of the multiple messages over the fronthaul links, incurring inter-message interference both at the cloud and ENs. To handle this issue, we propose efficient fronthaul multiple accessing schemes that design the uplink and downlink signaling strategies $u(\cdot)$ in \eqref{eq:y0} and $d_{i}(\cdot)$ in \eqref{eq:d}, respectively.

%

\subsection{OMA fronthauling}
We first develop an OMA method where distinct fronthaul resources are assigned to each of uplink and downlink messages to avoid inter-message interferences.
The uplink messages $\mathbf{m}_{i0}\in\mathbb{R}^{M_{i0}}$ for $i=1,\cdots,N$ occupy $N$ bundles of the fronthaul RBs where the $i$-th resource bundle containing $M_{i0}$ RBs is dedicated to the uplink message transmission of EN $i$. In this setup, the uplink signaling strategy $u(\cdot)$ in \eqref{eq:y0} becomes the concatenation operation. Then, the received signals at the cloud \eqref{eq:y0} is rewritten by
\begin{align}
\mathbf{y}_{0}^{\text{OMA}}=\bigoplus_{i=1}^{N}h_{i0}(\mathbf{m}_{i0}), \label{eq:y0_OMA}
\end{align}
where $\bigoplus_{i=1}^{N}\mathbf{q}_{i}\triangleq[\mathbf{q}_{1}^{T},\cdots,\mathbf{q}_{N}^{T}]^{T}$ defines the concatenation of $N$ vectors $\mathbf{q}_{i}$ for $i=1,\cdots,N$. The dimension of $\mathbf{y}_{0}^{\text{OMA}}$ becomes $M_{U}\triangleq\sum_{i=1}^{N}M_{i0}$ where $M_{U}$ indicates the total number of the uplink fronthaul RBs.

In the downlink, $\mathbf{m}_{0i}\in\mathbb{R}^{M_{0i}}$ is sent on $N$ orthogonal downlink fronthaul links each having $M_{0i}$ RBs. Hence, the downlink signaling strategy $d_{i}(\cdot)$ in \eqref{eq:d} can be specified as a masking operation extracting $\mathbf{m}_{0i}$ from the downlink message set $\{\mathbf{m}_{0j}:\forall j\}$, i.e., $\mathbf{d}_{i}=\mathbf{m}_{0i}$. Combining this with $\mathcal{M}_{0}(\cdot)$ in \eqref{eq:m0i}, the downlink message generation of the OMA system can be refined as the procedure that creates the concatenation of $N$ downlink messages. It follows
\begin{align}
\bigoplus_{i=1}^{N}\mathbf{m}_{0i}=[\mathbf{m}_{01}^{T},\cdots,\mathbf{m}_{0N}^{T}]^{T}=\mathcal{M}_{0}(\mathbf{y}_{0}^{\text{OMA}}).\label{eq:m0i_OMA}
\end{align}
The downlink message $\mathbf{m}_{0i}$ is then received by EN $i$ through the corresponding downlink fronthaul channel $h_{0i}(\cdot)$. Hence, we refine the received signal at EN $i$ in \eqref{eq:yi} as
\begin{align}
\mathbf{y}_{i}^{\text{OMA}}=h_{0i}(\mathbf{m}_{0i}). \label{eq:yi_OMA}
\end{align}
Since $M_{0i}$ RBs are allocated for the transmission of $\mathbf{m}_{0i}$, the length of $\mathbf{y}_{i}^{\text{OMA}}$ is given by $M_{0i}$, resulting in $M_{D}\triangleq\sum_{i=1}^{N}M_{0i}$ downlink fronthaul RBs. Therefore, the total number of the RBs denoted by $M$ is written by $M=M_{U}+M_{D}=\sum_{i=1}^{N}(M_{i0}+M_{0i})$.

\begin{rem}
The orthogonal interaction concept has been adopted in various decentralized optimization techniques such as the message-passing algorithms \cite{JLiu:17,FRKschi:01}, the ADMM framework \cite{YXiao:18,Boyd:10}, and the distributed learning systems \cite{DGunduz:19,Kim2018,HLee:19b,Yasar:19}. However, they do not consider the effect of the practical fronthaul links including the channel imperfection and the signaling overheads. Also, the effectiveness of the OMA interaction policy is not clearly addressed in the DNN-based optimization approaches \cite{DGunduz:19,Kim2018,HLee:19b,Yasar:19}. In the proceeding sections, we investigate the optimality of the proposed CECIL approach implemented with the OMA fronthauling scheme.
\end{rem}

\subsection{NOMA fronthauling}\label{sec:sec5a}
The OMA strategy may waste the fronthaul resources for allocating distinct RBs for each EN. To this end, we propose a non-orthogonal message transmission scheme where all ENs share the same fronthaul resources. Provided that $M_{U}$ RBs are assigned for the uplink message transmission, EN $i$ obtains its messages $\mathbf{m}_{i0}$ from \eqref{eq:mi0} by setting $M_{i0}=M_{U}$, i.e., utilizing all uplink fronthaul RBs. Then, the uplink transmission strategy $u(\cdot)$ is obtained as the superposition of all the downlink messages since they are interfere with each other. Therefore, the cloud receives the superposed signal $\mathbf{y}_{0}^{\text{NOMA}}\in\mathbb{R}^{M_{U}}$ of length $M_{U}$ expressed as
\begin{align}
\mathbf{y}_{0}^{\text{NOMA}}=\sum_{i=1}^{N}h_{i0}(\mathbf{m}_{i0}).
\end{align}

In the downlink, the cloud multicasts a common downlink message $\mathbf{m}_{0}\in\mathbb{R}^{M_{D}}$ of length $M_{D}$ to all the ENs by leveraging all the available $M_{D}$ downlink fronthaul RBs. Then, the downlink signaling in \eqref{eq:y0} is simply fixed as $\mathbf{d}_{i}=\mathbf{m}_{0}$, $\forall i$, such that the cloud directly transmits the output of the cloud computation in \eqref{eq:m0_NOMA}. We thus modify \eqref{eq:m0i} for the NOMA scheme as
\begin{align}
\mathbf{m}_{0}=\mathcal{M}_{0}(\mathbf{y}_{0}^{\text{NOMA}}). \label{eq:m0_NOMA}
\end{align}
Accordingly, the received signal $\mathbf{y}_{i}^{\text{NOMA}}\in\mathbb{R}^{M_{D}}$ of length $M_{D}$ at EN $i$ can be rewritten by
\begin{align}
\mathbf{y}_{i}^{\text{NOMA}}=h_{0i}(\mathbf{m}_{0}).
\end{align}

\subsection{Discussions}\label{sec:sec5C}
We discuss the effectiveness of the OMA and NOMA schemes for the perfect fronthaul link case, i.e., $h_{i0}(\cdot)$ and $h_{0i}(\cdot)$ are given by the identity functions. The received signals of the OMA and NOMA systems are respectively recast to
\begin{align}
\mathbf{y}_{0}^{\text{OMA}}=\bigoplus_{i=1}^{N}\mathbf{m}_{i0},\ \mathbf{y}_{i}^{\text{OMA}}=\mathbf{m}_{0i},\label{eq:perfectOMA}\\
\mathbf{y}_{0}^{\text{NOMA}}=\sum_{i=1}^{N}\mathbf{m}_{i0},\ \mathbf{y}_{i}^{\text{NOMA}}=\mathbf{m}_{0},\label{eq:perfectNOMA}
\end{align}
which simplifies \eqref{eq:m0i_OMA} and \eqref{eq:m0_NOMA} as
\begin{align}
\bigoplus_{i=1}^{N}\mathbf{m}_{0i}&=\mathcal{M}_{0}\left(\bigoplus_{i=1}^{N}\mathcal{M}_{i}(\mathbf{a}_{i})\right), \label{eq:m0i_OMA_perfect}\\
\mathbf{m}_{0}&=\mathcal{M}_{0}\left(\sum_{i=1}^{N}\mathcal{M}_{i}(\mathbf{a}_{i})\right). \label{eq:m0_NOMA_perfect}
\end{align}

\subsubsection{Optimality of NOMA fronthauling}
We first focus on the NOMA system.
For constructing successful decision inference $\mathcal{X}_{i}(\cdot)$ in \eqref{eq:xi}, the optimal downlink message denoted by $\mathbf{m}_{0}^{\star}$ needs to properly encode all local observations $\mathbf{a}_{i}$, $\forall i$. Also, since the NOMA downlink message $\mathbf{m}_{0}^{\star}$ is common for all ENs, it should not be affected by permutations of input features. In other words, the computation of the downlink message has to be independent of the ordering of $\mathbf{a}_{i}$, $\forall i$, so that individual ENs can leverage the downlink message for the individual decision $\mathbf{x}_{i}=\mathcal{X}_{i}(\mathbf{a}_{i},\mathbf{m}_{0})$ without knowing their order $i$ indexed by the network. Notice that such a permutation-invariant property indeed holds for \eqref{eq:m0_NOMA_perfect} due to the superposition signaling in \eqref{eq:perfectNOMA}.

Based on this intuition, we can model the optimal downlink message $\mathbf{m}_{0}^{\star}$ of the NOMA by using a generic set operator $g(\cdot)$, which is defined over a set of the local observations $\{\mathbf{a}_{i}:\forall i\}$, to satisfy the permutation-invariant property. The corresponding formulation can be written as
\begin{align}
\mathbf{m}_{0}^{\star}=g(\{\mathbf{a}_{i}:\forall i\}). \label{eq:m0star}
\end{align}
It is easy to see that \eqref{eq:m0star} does not change with the ordering of the ENs since the input feature is given by the set. We may lose the optimality in the NOMA system if the downlink message calculation strategy \eqref{eq:m0_NOMA_perfect} cannot approximate the optimal one in \eqref{eq:m0star} accurately. The following proposition states that \eqref{eq:m0_NOMA_perfect} can be the universal approximator for an arbitrary set function.
\begin{prop}\label{prop:prop1}
Suppose that the local observation $\mathbf{a}_{i}$ is drawn from a compact set $\mathcal{A}_{i}$ and has the identical dimension. Let $g(\cdot)$ be any continuous set function with the permutation-invariant property that maps $N$ local observations to $M_{D}$-dimensional output vector. Then, for arbitrary small $\varepsilon>0$, there exist an outer mapping $\mathcal{M}_{0}(\cdot)$ and an inner mapping $\mathcal{M}_{i}(\cdot)$ satisfying
\begin{align}
\sup_{\{\mathbf{a}_{i}\in\mathcal{A}_{i},\forall i\}}\left\|g(\{\mathbf{a}_{i}:\forall i\})-\mathcal{M}_{0}\left(\sum_{i=1}^{N}\mathcal{M}_{i}(\mathbf{a}_{i})\right)\right\|<\varepsilon. \label{eq:ep}
\end{align}
\end{prop}
\begin{IEEEproof}
Let $[\mathbf{u}]_{k}$ be the $k$-th element of a vector $\mathbf{u}$. Suppose an arbitrary set function $\lambda(\{\mathbf{a}_{i}:\forall i\})$ whose output is given by a scalar number. From \cite[Thm. 9]{MZaheer:17} and the Stone–Weierstrass theorem \cite{Cotter:90}, there exist a continuous mapping $m_{k}:\mathbb{R}^{M_{U}}\rightarrow\mathbb{R}$ and arbitrary small $\varepsilon_{k}>0$ which fulfills
\begin{align}
\sup_{\{\mathbf{a}_{i}\in\mathcal{A}_{i},\forall i\}}\left|\lambda(\{\mathbf{a}_{i}:\forall i\})\!\!-\!m_{k}\left(\sum_{i=1}^{N}\mathcal{M}_{i}(\mathbf{a}_{i})\right)\right|\!<\!\varepsilon_{k}. \label{eq:epk}
\end{align}
By setting $\lambda(\{\mathbf{a}_{i}:\forall i\})=[g(\{\mathbf{a}_{i}:\forall i\})]_{k}$ in \eqref{eq:epk}, it is concluded $m_{k}(\cdot)$ forms the universal approximator for the $k$-th element of the optimal message vector $[\mathbf{m}_{0}^{\star}]_{k}=[g(\{\mathbf{a}_{i}:\forall i\})]_{k}$.
Stacking $M_{D}$ element-wise mappings $m_{k}(\cdot)$ for $k=1,\cdots,M_{D}$ leads to \eqref{eq:ep} with $\mathcal{M}_{0}(\cdot)=\bigoplus_{k=1}^{M_{D}}m_{k}(\cdot)$ and $\varepsilon_{k}=\frac{\varepsilon}{\sqrt{M_{D}}}$. This completes the proof.
\end{IEEEproof}

Notice that the optimal downlink message generation \eqref{eq:m0star} cannot be implemented in the practical F-RAN systems since the cloud needs to know the local statistics of the ENs perfectly. Nevertheless, thanks to Proposition \ref{prop:prop1}, it can be alternatively executed through the proposed computation rule in \eqref{eq:m0_NOMA_perfect}. Thus, although the uplink messages are independently created at the ENs, the superposition signaling and resource sharing policies of the uplink NOMA fronthauling strategy leads to the successful distributed decision at the ENs. Since Proposition \ref{prop:prop1} holds for any continuous functions $\mathcal{M}_{0}(\cdot)$ and $\mathcal{M}_{i}(\cdot)$, the universal approximation property is satisfied in the DL formulation with well-designed DNNs \eqref{eq:mi0_fnn} and \eqref{eq:m0i_fnn}. As a result, the unknown optimal downlink message $\mathbf{m}_{0}^{\star}$ can be obtained by optimizing the DNNs with the end-to-end training policy \eqref{eq:sgd}.

\subsubsection{Impact of $M_{U}$}\label{sec:sec5C2}
We analyze the number of the uplink fronthaul RBs $M_{U}$ required for achieving the universal approximation property \eqref{eq:ep}. For a scalar input $u$, a simple inner mapping $\mathcal{M}_{i}(u)=[1,u,u^2,\cdots,u^{N}]^{T}$ of length $N+1$ achieves the element-wise universal approximation property \eqref{eq:epk} \cite[Thm. 7]{MZaheer:17}. This implies that $M_{U}=N+1$ uplink fronthaul RBs are sufficient if all the local observations $\mathbf{a}_{i}$, $\forall i$, are given by scalar numbers. An extension to a general vector input case is challenging. Instead, we may consider a trivial modification of \eqref{eq:m0star} as
\begin{align}
\mathbf{m}_{0}^{\star}=g(\{\mathbf{a}_{i}:\forall i\})=g(\{[\mathbf{a}_{i}]_{l}:\forall i, l=1,\cdots,A_{i}\}), \label{eq:m0star2}
\end{align}
where the observation vector $\mathbf{a}_{i}\in\mathbb{R}^{A_{i}}$ is decoupled into its $A_{i}$ elements $[\mathbf{a}_{i}]_{l}$ for $l=1,\cdots,A_{i}$. A modified operator now converts a set of $\sum_{i=1}^{N}A_{i}$ elements into $M_{D}$-dimensional downlink message vector. This preserves the optimality since the resulting message still involves the global state $\mathbf{a}=\{\mathbf{a}_{i}:\forall i\}$ essential for the individual decision of the ENs. To implement \eqref{eq:m0star2}, EN $i$ can employ $A_{i}$ different operators $\mathcal{M}_{il}([\mathbf{a}_{i}]_{l})=[1,[\mathbf{a}_{i}]_{l},[\mathbf{a}_{i}]_{l}^2,\cdots,[\mathbf{a}_{i}]_{l}^{A}]^{T}$, $\forall l=1,\cdots,A_{i}$. Then, \eqref{eq:m0_NOMA_perfect} can be recast to
\begin{align}
\mathbf{m}_{0}=\mathcal{M}_{0}\left(\sum_{i=1}^{N}\sum_{l=1}^{A_{i}}\mathcal{M}_{il}([\mathbf{a}_{i}]_{l})\right). \label{eq:m0_NOMA_perfect2}
\end{align}

The NOMA strategy in \eqref{eq:m0_NOMA_perfect2} is achieved with $M_{U}=\sum_{i=1}^{N}A_{i}+1$ uplink fronthaul RBs. Although \eqref{eq:m0_NOMA_perfect2} is proven to be effective, we adopt the vector-valued operator $\mathcal{M}_{i}:\mathbb{R}^{A_{i}}\rightarrow \mathbb{R}^{M_{U}}$ as in \eqref{eq:m0_NOMA_perfect} since it includes \eqref{eq:m0_NOMA_perfect2} as a special case by restricting weight matrices of the DNN in \eqref{eq:mi0_fnn} to diagonal matrices. Numerical results confirm that \eqref{eq:m0_NOMA_perfect} requests a much smaller number of the uplink fronthaul RBs than the analytical result $M_{U}=\sum_{i=1}^{N}A_{i}+1$.

\subsubsection{Optimality of OMA fronthauling}
We now discuss the optimality of the OMA scheme in \eqref{eq:m0i_OMA_perfect}. Thanks to the orthogonal transmission, the cloud can separate the uplink messages $\mathbf{m}_{i0}=\mathcal{M}_{i}(\mathbf{a}_{i})$, $\forall i$.
Nevertheless, the universal approximation theorem \eqref{eq:univ1} cannot be constructed for \eqref{eq:m0i_OMA_perfect} since a simple concatenation of DNNs $\bigoplus_{i=1}^{N}\mathcal{M}_{i}(\cdot)$ is far from the fully-connected DNN assumed in \eqref{eq:univ1}. To this end, we present a suitable transformation of \eqref{eq:m0i_OMA_perfect} that removes the concatenation operations. Let $\tilde{\mathbf{m}}_{0i}\triangleq\tilde{\mathcal{M}}_{i}(\mathbf{a}_{i})\in\mathbb{R}^{M_{U}}$ of length $M_{U}$ be a zero-padded version of $\mathbf{m}_{i0}\in\mathbb{R}^{M_{i0}}$. All elements of $\tilde{\mathbf{m}}_{0i}$ are zeros except the $(\sum_{j=1}^{i-1}M_{j0}+1)$-th to the $(\sum_{j=1}^{i}M_{j0})$-th elements being replaced with $\mathbf{m}_{i0}$. Similarly, the corresponding message generation operator $\tilde{\mathcal{M}}_{i}(\cdot)$ can also be defined as the zero-padded version of the original inference $\mathcal{M}_{i}(\cdot)$. Then, \eqref{eq:m0i_OMA_perfect} can be refined~as
\begin{align}
\tilde{\mathbf{m}}_{0}=\mathcal{M}_{0}\left(\sum_{i=1}^{N}\tilde{\mathcal{M}}_{i}(\mathbf{a}_{i})\right), \label{eq:m0i_OMA_perfect2}
\end{align}
where $\tilde{\mathbf{m}}_{0}\triangleq\bigoplus_{i=1}^{N}\mathbf{m}_{0i}\in\mathbb{R}^{M_{D}}$ is the concatenation of the downlink messages. Unlike the NOMA case \eqref{eq:m0_NOMA_perfect}, due to the concatenation operation, the ordering of the ENs affects the downlink message computations of the OMA. Therefore, the optimal OMA downlink message $\tilde{\mathbf{m}}_{0}^{\star}$ is modeled as a generic inference $\tilde{g}(\cdot)$ with the stacked local observation vectors, i.e., $\tilde{\mathbf{m}}_{0}^{\star}=\tilde{g}(\bigoplus_{i=1}^{N}\mathbf{a}_{i})$, rather than the permutation-invariant set function in \eqref{eq:m0star}. Proposition~\ref{prop:prop1}, which is based on the permutation-invariance of the target set function, cannot be straightforwardly applied to the OMA method.

To address this, we leverage the Kolmogorov–Arnold representation theorem \cite{VKarkova:92} which states that any continuous mapping can be represented as a superposition of continuous functions. Assuming $M_{D}=1$ and scalar local observations $a_{i}\in\mathbb{R}$, $\forall i$, a continuous function $\tilde{g}(\bigoplus_{i=1}^{N}a_{i})$ has the following representation \cite[Thm. 8]{MZaheer:17}
\begin{align}
\tilde{g}\left(\bigoplus_{i=1}^{N}a_{i}\right)=\mathcal{M}_{0}\left(\sum_{i=1}^{N}\tilde{\mathcal{M}}_{i}(a_{i})\right) \label{eq:arnold}
\end{align}
with some mappings $\mathcal{M}_{0}:\mathbb{R}^{2N+1}\rightarrow\mathbb{R}$ and $\tilde{\mathcal{M}}_{i}:\mathbb{R}\rightarrow\mathbb{R}^{2N+1}$. The uplink message generation operator of the OMA $\tilde{\mathcal{M}}_{i}(\cdot)$ requires $M_{U}=2N+1$ uplink fronthaul RBs for the universal approximation property. With similar approaches presented in Section \ref{sec:sec5C2}, extensions of \eqref{eq:arnold} to the general case with $M_{D}>1$ and vector inputs $\mathbf{a}_{i}\in\mathbb{R}^{A_{i}}$, $\forall i$, result in $M_{U}=2\sum_{i=1}^{N}A_{i}+1$ uplink RBs, which is about twice as large as that of the NOMA case in \eqref{eq:m0_NOMA_perfect2} achieved with $M_{U}=\sum_{i=1}^{N}A_{i}+1$. Thus, although the performance of the OMA method could reach that of the NOMA system, it might need more uplink fronthaul resources. This is verified from the numerical results.

\section{Imperfect Fronthaul Links}\label{sec:sec6}
This section investigates the imperfect fronthaul link cases with random noise and finite capacity constraints. The robust training strategy of the CECIL framework is proposed for each scenario. The details are explained in the following.

\subsection{Noisy fronthaul links}\label{sec:sec6A}
The imperfection of the wireless fronthauls can be modeled by the random additive noise. We specify the fronthaul channel functions as $h_{i0}(\mathbf{u})=h_{0i}(\mathbf{u})=\mathbf{u}+\boldsymbol{\eta}$ where $\boldsymbol{\eta}$ stands for the noise vector with arbitrary distribution. In the OMA system, the received messages $\mathbf{y}_{0}^{\text{OMA}}$ at the cloud \eqref{eq:y0_OMA} and $\mathbf{y}_{i}^{\text{OMA}}$ at EN $i$ \eqref{eq:yi_OMA} are respectively written by
\begin{align}
\mathbf{y}_{0}^{\text{OMA}}=\bigoplus_{i=1}^{N}\mathbf{m}_{i0}+\boldsymbol{\eta}_{0}\ \text{and}\ \mathbf{y}_{i}^{\text{OMA}}=\mathbf{m}_{0i}+\boldsymbol{\eta}_{i}, \label{eq:yOMA_AWGN}
\end{align}
where $\boldsymbol{\eta}_{i}$ for $i=0,1,\cdots,N$ denotes the noise at node $i$. We obtain similar formulations for the NOMA system as
\begin{align}
\mathbf{y}_{0}^{\text{NOMA}}=\sum_{i=1}^{N}\mathbf{m}_{i0}+\boldsymbol{\eta}_{0}\ \text{and}\ \mathbf{y}_{i}^{\text{NOMA}}=\mathbf{m}_{0}+\boldsymbol{\eta}_{i}. \label{eq:yNOMA_AWGN}
\end{align}
The noise hinders successful decisions at the ENs, thereby requiring robust message generation strategies both at the cloud and ENs. To this end, we modify the training update in \eqref{eq:sgd} by taking the noise into account. We include numerous realization of the random noise vectors into the training set. A mini-batch set $\mathcal{A}$ becomes a set of tuples $(\mathbf{a},\{\boldsymbol{\eta}_{i}:\forall i\})$ of the global observation $\mathbf{a}$ and a collection of the uplink and downlink noise vectors $\{\boldsymbol{\eta}_{i}:\forall i\}$.
The DNN parameter $\boldsymbol{\Theta}$ is then adjusted in the ascent direction of the gradient averaged over the noise distribution.
Such a data-driven optimization enables robust design of the CECIL by observing numerous noisy messages \eqref{eq:yOMA_AWGN} and \eqref{eq:yNOMA_AWGN} in the training step.

\subsection{Finite-capacity fronthaul links}\label{sec:sec6B}
Until Section. \ref{sec:sec5}, we assumed lossless fronthaul interactions where each RB can convey a real-valued scalar number without any distortion. In the practical wired fronthaul link setup, however, the resolution of the message would be limited by the fronthaul capacity. To this end, in this subsection, we design a robust training policy of the CECIL for the general case where the fronthaul links are subject to the transmission capacity. The fronthaul channels $h_{i0}(\cdot)$ and $h_{0i}(\cdot)$ can be given as the rounding functions that output the nearest integer of the transmitted messages. In this configuration, only the lossy coordination is allowed to share discrete-valued messages. To accommodate capacity-limited fronthaul links, we present a message quantization process that creates discrete representations of continuous-valued messages. We focus on the quantization of the uplink message $\mathbf{m}_{i0}$, but the proposed techniques are readily applied to the downlink message quantization. Let $\hat{\mathbf{m}}_{i0}\in\mathbb{R}^{M_{i0}}$ be the quantization output of $\mathbf{m}_{i0}$. The capacity of the uplink fronthaul link connecting EN $i$ and the cloud is modeled by a set of integers $C_{il}$, $\forall l=1,\cdots,M_{i0}$, each of which indicates the alphabet size, or equivalently, the modulation level allowed for transferring the $l$-th element $[\mathbf{m}_{i0}]_{l}$. It is expressed as
\begin{align}
[\hat{\mathbf{m}}_{i0}]_{l}\triangleq\psi_{C_{il}}([\mathbf{m}_{i0}]_{l})\in\{0,1,\cdots,C_{il}-1\},\label{eq:mhat}
\end{align}
where $\psi_{C}(\cdot)$ stands for the quantization function with the quantization level $C$. It maps a continuous-valued input into a discrete set $\{0,1,\cdots,C-1\}$. The received signals in \eqref{eq:y0} and \eqref{eq:yi} can then be refined as $\mathbf{y}_{0}=u\big(\{h_{i0}(\hat{\mathbf{m}}_{i0}):\forall i\}\big)$ and $\mathbf{y}_{i}=h_{0i}\big(d_{i}(\{\hat{\mathbf{m}}_{0j}:\forall j\})\big)$, respectively.

The quantization operator $\psi_{C_{il}}(\cdot)$ is viewed as an activation function that is followed by $\mathcal{M}_{i}(\cdot)$, i.e., the DNN $\mathcal{F}_{Q_{M_{i}}}(\cdot;\theta_{M_{i}})$ in \eqref{eq:mi0_fnn}. Our target is to design the activation $\psi_{C_{il}}(\cdot)$ such that $\hat{\mathbf{m}}_{i0}$ acts as an accurate estimate of the original message $\mathbf{m}_{i0}$. In this way, the cloud and ENs can successfully recover the original messages through their quantized observations. One naive approach would be to employ the rounding function. However, the simple rounding activation exhibits zero gradient for all input regime, thereby prohibiting the DNN parameters from being optimized using the SGD method in \eqref{eq:sgd}. This has been well-known as the vanishing gradient problem where the performance of the DNNs are no longer improved but possibly gets stuck into an unsatisfactory point \cite{IGoodfellow:16}. In our case, the DNNs $\mathcal{F}_{Q_{M_{i}}}(\cdot;\theta_{M_{i}})$ in \eqref{eq:mi0_fnn} and \eqref{eq:m0i_fnn} would not be trained properly. To handle this difficulty, a novel quantization method has been provided in \cite{HLee:19b,HLee:20}, but it is only applicable to the special case of $C_{il}=2$.

We propose an integerization technique which is regarded as an extension of the binarization method in \cite{HLee:19b} for the general case of $C_{il}>2$. The $l$-th element of the continuous-valued message $\mathbf{m}_{i0}$ is assumed to lie in a bounded region $[0,C_{il}-1]$. This can be achieved by applying a bounding activation, e.g., the sigmoid function, to the output layer of the DNN $\mathcal{F}_{Q_{M_{i}}}(\cdot;\theta_{M_{i}})$. The proposed quantization function $\psi_{C_{il}}(\cdot)$ in \eqref{eq:mhat} carries out a randomized rounding operation. It first configures two nearest integers $c-1$ and $c$, $\forall c=1,\cdots,C_{il}-1$, of the input $[\mathbf{m}_{i0}]_{l}$, i.e., $[\mathbf{m}_{i0}]_{l}\in[c-1,c)$, as candidates of the quantization. For notational simplicity, we denote $m\triangleq[\mathbf{m}_{i0}]_{l}$ and $\hat{m}\triangleq[\hat{\mathbf{m}}_{i0}]_{l}$. Provided that $m\in[c-1,c)$, the rounding output $\hat{m}=\psi_{C_{il}}(m)$ can be either $c-1$ or $c$ with probabilities
\begin{align}
\Pr\{\hat{m}&=c-1|m\in[c-1,c)\}=(c-m), \label{eq:pc1}\\
\Pr\{\hat{m}&=c|m\in[c-1,c)\}=(m-(c-1)). \label{eq:pc}
\end{align}
The probabilities in \eqref{eq:pc1} and \eqref{eq:pc} can be interpreted as the distances from the continuous input $m$ to the target quantization points $c$ and $c-1$, respectively. The probability $\Pr\{\hat{m}=c|m\in[c-1,c)\}$ increases as $m$ gets closer to $c$, and the resulting quantization $\hat{m}$ is more likely to be $c$.

The proposed quantization activation $\psi_{C_{il}}(m)$ for an input $m\in[0,C_{il}-1)$ is given as
\begin{align}
    \!\!\!\!\!\psi_{C_{il}}(m)\!=\!\!\begin{cases}\label{eq:psi}
    c-1, &\!\!\!\!\!  \text{with prob.}\ (c-m)\!\cdot\!\mathds{1}_{m\in[c-1,c)},\\
    c, &\!\!\!\!\! \text{with prob.}\ (m-(c-1))\!\cdot\!\mathds{1}_{m\in[c-1,c)},
  \end{cases}
\end{align}
where $\mathds{1}_{Z}\in\{0,1\}$ denotes the indicator function which is $1$ if the condition $Z$ is true and $0$ otherwise. The following proposition states the quality of the quantization $\hat{m}=\psi_{C_{il}}(m)$ in terms of its estimation property for unavailable information~$m$.
\begin{prop}\label{prop:prop2}
The quantization $\hat{m}=\psi_{C_{il}}(m)$ with the probabilities \eqref{eq:pc1} and \eqref{eq:pc} is an unbiased estimate of $m$.
\end{prop}
\begin{IEEEproof}
To prove the unbiased estimation property, it suffices to show that the conditional expectation of $\hat{m}$ given $m$, denoted by $\mathbb{E}_{\hat{m}}[\hat{m}|m]$, is equal to $m$. It follows
\begin{align}
\mathbb{E}_{\hat{m}}[\hat{m}|m]&=\mathbb{E}_{c}\big[\mathbb{E}_{\hat{m}}[\hat{m}|m\in[c-1,c)]\big]\\
    &=\sum_{c=1}^{C_{il}}\!\Pr\{m\!\in\![c\!-\!1,c)\}\mathbb{E}_{\hat{m}}[\hat{m}|m\!\in\![c\!-\!1,c)]\\
    &=\sum_{c=1}^{C_{il}}\Pr\{m\in[c-1,c)\}\cdot m=m, \label{eq:unbiased}
\end{align}
where \eqref{eq:unbiased} is obtained since
\begin{align}
\mathbb{E}_{\hat{m}}[\hat{m}|m\in[c-1,c)]
    &=(c-1)\cdot\Pr\{\hat{m}=c-1|m\in[c-1,c)\}\\
    &~~~~+c\cdot\Pr\{\hat{m}=c|m\in[c-1,c)\}=m
\end{align}
We thus have $\mathbb{E}_{\hat{m}}[\hat{m}|m]=m$. This completes the proof.
\end{IEEEproof}
Proposition \ref{prop:prop2} reveals that the cloud and ENs can accurately recover the continuous-valued messages by taking expectations over the received quantized messages. This can be realized with numerous quantization samples observed in the training step. Therefore, the DNN at the cloud $\mathcal{F}_{Q_{M_{0}}}(\cdot;\theta_{M_{0}})$ in \eqref{eq:m0i_fnn}, which processes the quantized uplink messages $\hat{\mathbf{m}}_{i0}$, can be trained to decode the original information $\mathbf{m}_{i0}$ successfully. 

Now, we discuss an efficient training strategy of the DNNs implemented with the probabilistic activation \eqref{eq:psi}, which is, in general, has no closed-form expression for the gradient $\nabla_{\boldsymbol{\Theta}}\psi_{C_{il}}(m)$. To address this, the gradient estimation techniques \cite{YBengio:13,Raiko:15,HLee:19b,HLee:20} is adopted which approximate an intractable gradient with its average evaluated over any randomized operations. By leveraging Proposition \ref{prop:prop2}, the gradient $\nabla_{\boldsymbol{\Theta}}\psi_{C_{il}}(m)$ can be approximated as
\begin{align}
\nabla_{\boldsymbol{\Theta}}\psi_{C_{il}}(m)=\nabla_{\boldsymbol{\Theta}}\hat{m}\simeq\nabla_{\boldsymbol{\Theta}}\mathbb{E}_{\hat{m}}[\hat{m}|m]=\nabla_{\boldsymbol{\Theta}}m.\label{eq:gradm}
\end{align}
It is inferred from \eqref{eq:gradm} that the gradient of the proposed quantization activation can be simply replaced with that of the input continuous-valued message $m$. Since $m$ is obtained with a bounding activation, e.g., sigmoid function, whose derivative is well-defined in all the input domain, the parameter set $\boldsymbol{\Theta}$ can be efficiently trained with the SGD algorithm.

Combining \eqref{eq:psi} and \eqref{eq:gradm}, we can conclude that the proposed quantization activation exhibits different behaviors in the forward pass and backward pass. The actual quantized messages are computed in the forward pass with the randomized rounding operations \eqref{eq:psi}, and the resulting quantization is forwarded through the capacity-limited fronthaul links. On the contrary, to optimize the DNN parameter set $\boldsymbol{\Theta}$, we need to calculate the gradients through the backpropagation algorithm \cite{IGoodfellow:16}. In this backward pass computation, the quantization activation $\psi_{C_{il}}(\cdot)$ yields its input variable $m$ directly.


\section{Performance Evaluation}\label{sec:sec7}
This section assesses the performance of the proposed CECIL framework for power control applications in the F-RAN systems. 
EN $i$ ($i=1,\cdots,N$) sends data symbols to its intended mobile receiver referred to as user $i$. The ENs share the identical time-frequency resources for the data transmission. To mitigate the multi-user interference, an appropriate power allocation mechanism is required at individual ENs. The decision variable of EN $i$ becomes the transmit power $x_{i}\in[0,P]$ with $P$ equal to the maximum allowable power budget. Let $a_{ji}$ ($i,j=1,\cdots,N$) be the channel gain from EN $j$ to user $i$. EN $i$ can only observe an $N$-dimensional local CSI vector $\mathbf{a}_{i}\triangleq\{a_{ji}:\forall j\}\in\mathbb{R}^{N}$ that is reported from the corresponding user \cite{WLee:18a,HLee:19b}. The global network CSI is then defined as $\mathbf{a}=\{\mathbf{a}_{i}:\forall i\}\in\mathbb{R}^{N^{2}}$.

Two different utility functions are considered: average sum rate utility and average sum energy-efficiency (EE) utility. Defining $\mathcal{D}\triangleq\{\mathbf{x}|x_{i}\in[0,P]:\forall i\}$ as the feasible set of the concatenated solution vector $\mathbf{x}=\{x_{i}:\forall i\}\in\mathbb{R}^{N}$, the sum rate maximization (SRMax) and the sum EE maximization (EEMax) problems are respectively formulated as
\begin{align}
\max_{\mathbf{x}\in\mathcal{D}}\mathbb{E}_{\mathbf{a}}\bigg[\sum_{i=1}^{N}r_{i}(\mathbf{a},\mathbf{x})\bigg]\ \text{and}\ \max_{\mathbf{x}\in\mathcal{D}}\mathbb{E}_{\mathbf{a}}\bigg[\sum_{i=1}^{N}\frac{r_{i}(\mathbf{a},\mathbf{x})}{x_{i}+P_{S}}\bigg],\label{eq:app}
\end{align}
where $r_{i}(\mathbf{a},\mathbf{x})\triangleq\log(1+\frac{a_{ii}x_{i}}{1+\sum_{j\neq i}a_{ji}x_{j}})$ stands for the rate of user $i$ and $P_{S}$ is the static power consumption at ENs \cite{CIsheden:11}. The power of the proposed DL-based cooperative mechanisms and our intuitions presented in Section \ref{sec:sec5} can be analyzed by power control problems in \eqref{eq:app} which have been popular applications of DNN-assisted cooperative optimization studies \cite{PKerret:18,HLee:19b,Kim2018,Yasar:19}.

The channel gains are generated as the exponential random variables with unit mean. The transmit power constraint is set to $P=10$, and the static power consumption is fixed as $P_{S}=1$. A five-layer DNN with 100 hidden neurons is employed at the cloud DNN in \eqref{eq:m0i_fnn}. The DNNs \eqref{eq:mi0_fnn} and \eqref{eq:xi_fnn} at the ENs are constructed with three layers each with 50 neurons. The batch normalization technique \cite{Ioffe:15} followed by the rectified linear unit (ReLU) activation is adopted at the hidden layers. Unless stated otherwise, we use the linear activations at the output layers of the message generating DNNs at the cloud \eqref{eq:m0i_fnn} and at the ENs \eqref{eq:mi0_fnn}. For creating a feasible power level $x_{i}\in[0,P]$, the sigmoid function multiplied by $P$ is utilized at the output layer of the distributed optimizing DNN in \eqref{eq:xi_fnn}. Each training epoch consists of 50 mini-batches each of which contains $5000$ independently generated random channel gains $\mathbf{a}$. The Adam algorithm \cite{Kingma:15} with learning rate $\alpha=0.0001$ is exploited. The test performance is evaluated with $10^4$ test samples. The training and testing steps are implemented with Tensorflow 1.15.0 on a PC with an Intel i7-9700K CPU, 32 GB of RAM, and a GEFORCE RTX 2080 GPU.

\subsection{Perfect Fronthaul Link Case}\label{sec:sec7A}
We first focus on the perfect fronthaul link case where the messages can be exchanged via the noiseless fronthaul channels \eqref{eq:perfectOMA} and \eqref{eq:perfectNOMA}. In this ideal scenario, we validate the optimality of the NOMA and OMA fronthauling methods. The following baseline schemes are considered.
\begin{itemize}
\item {\em Ideal cooperation (IC):} The cloud is assumed to get the global CSI vector $\mathbf{a}$ perfectly. The cloud centrally computes the solution $\mathbf{x}$ via a DNN with 12 layers and 100 hidden neurons, which has the similar number of trainable variables to the proposed CECIL. The resulting solution is then assumed to be perfectly known to the ENs.
\item {\em No cooperation (NC):} No message exchange is allowed. Each EN needs to decide the power control solution with an individual DNN, which accepts only the local CSI $\mathbf{a}_{i}$ as input.
\item {\em Projected gradient descent (PGD):} The power control solution is optimized via the PGD method \cite{Boyd:04} under the feasible set $x_{i}\in[0,P]$. To facilitate GPU-enabled parallel computations, we utilize the Adam optimizer in Tensorflow with the precision $10^{-5}$. The PGD generates a locally optimum solution for the SRMax and EEMax.
\end{itemize}
To implement the IC and PGD methods, EN $i$ uploads an $N$-dimensional local CSI vector $\mathbf{a}_{i}$ to the cloud by using $M_{i0}=N$ RBs, resulting in total $M_{U}=\sum_{i=1}^{N}M_{i0}=N^{2}$ uplink fronthaul RBs. Also, the clouds forwards the local decision variable $x_{i}$ to EN $i$ through the downlink fronthaul links with $M_{0i}=1$ RB, requiring $M_{D}=\sum_{i=1}^{N}M_{0i}=N$ downlink RBs. Therefore, the total number of the fronthaul RBs is given as $M=M_{U}+M_{D}=N(N+1)$. On the other hands, the NC baseline does not allow no interactions among the cloud and the ENs as $M=0$.

\begin{figure*}
\centering
    \subfigure[$N=5$]{
        \includegraphics[width=.33\linewidth]{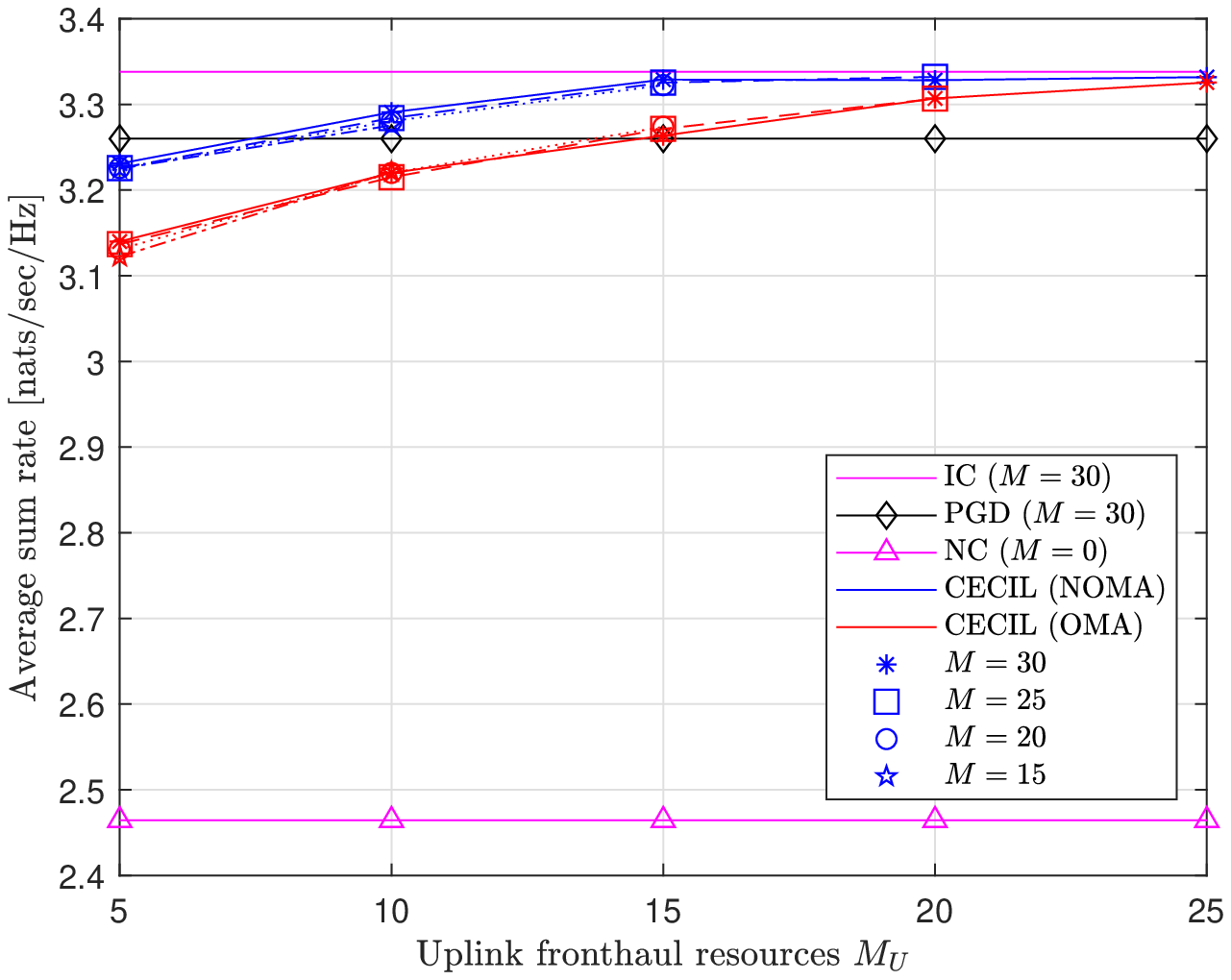}\label{fig:fig4a}
    }\hspace{-5mm}
    \subfigure[$N=7$]{
        \includegraphics[width=.33\linewidth]{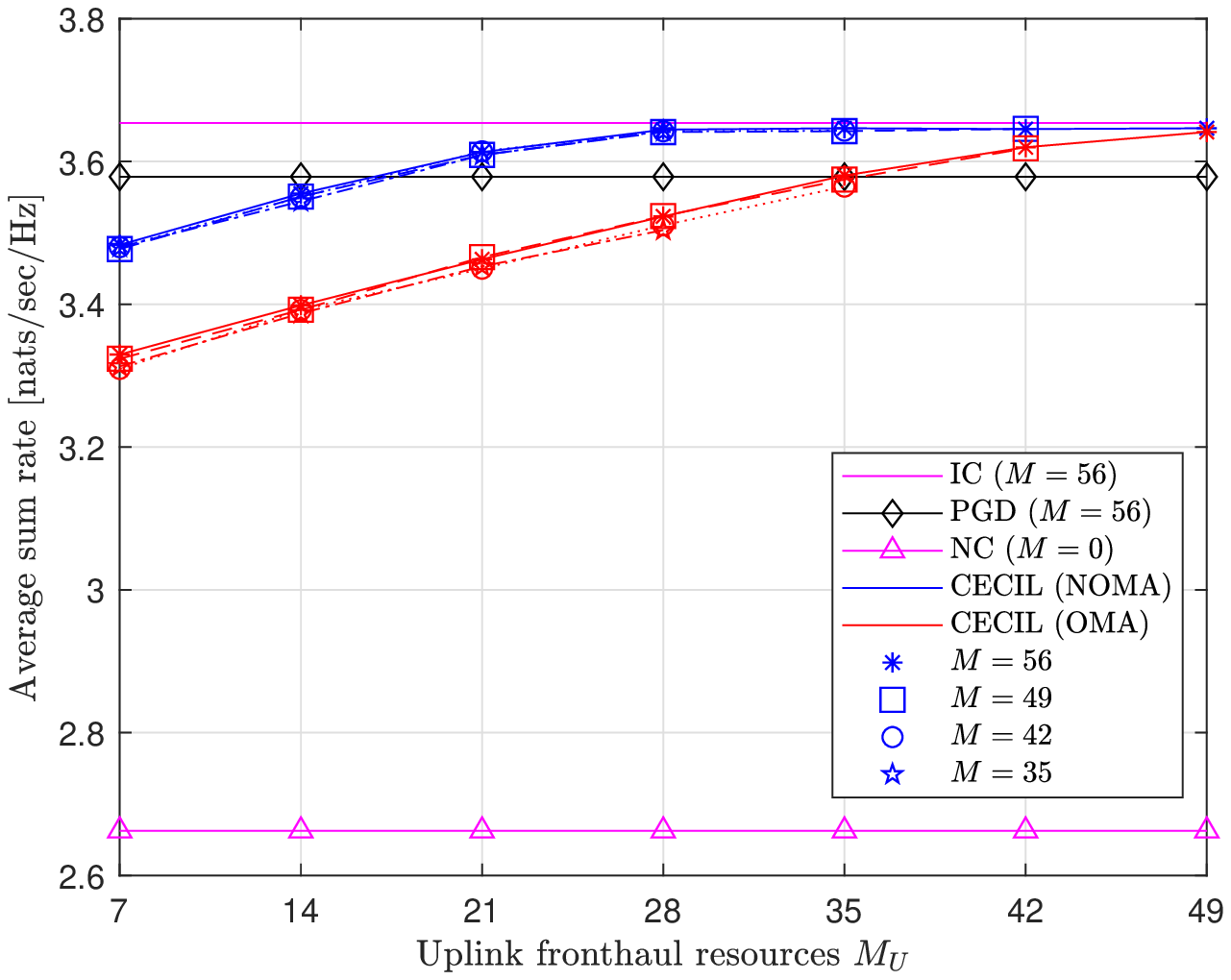}\label{fig:fig4b}
    }\hspace{-5mm}
    \subfigure[$N=9$]{
        \includegraphics[width=.33\linewidth]{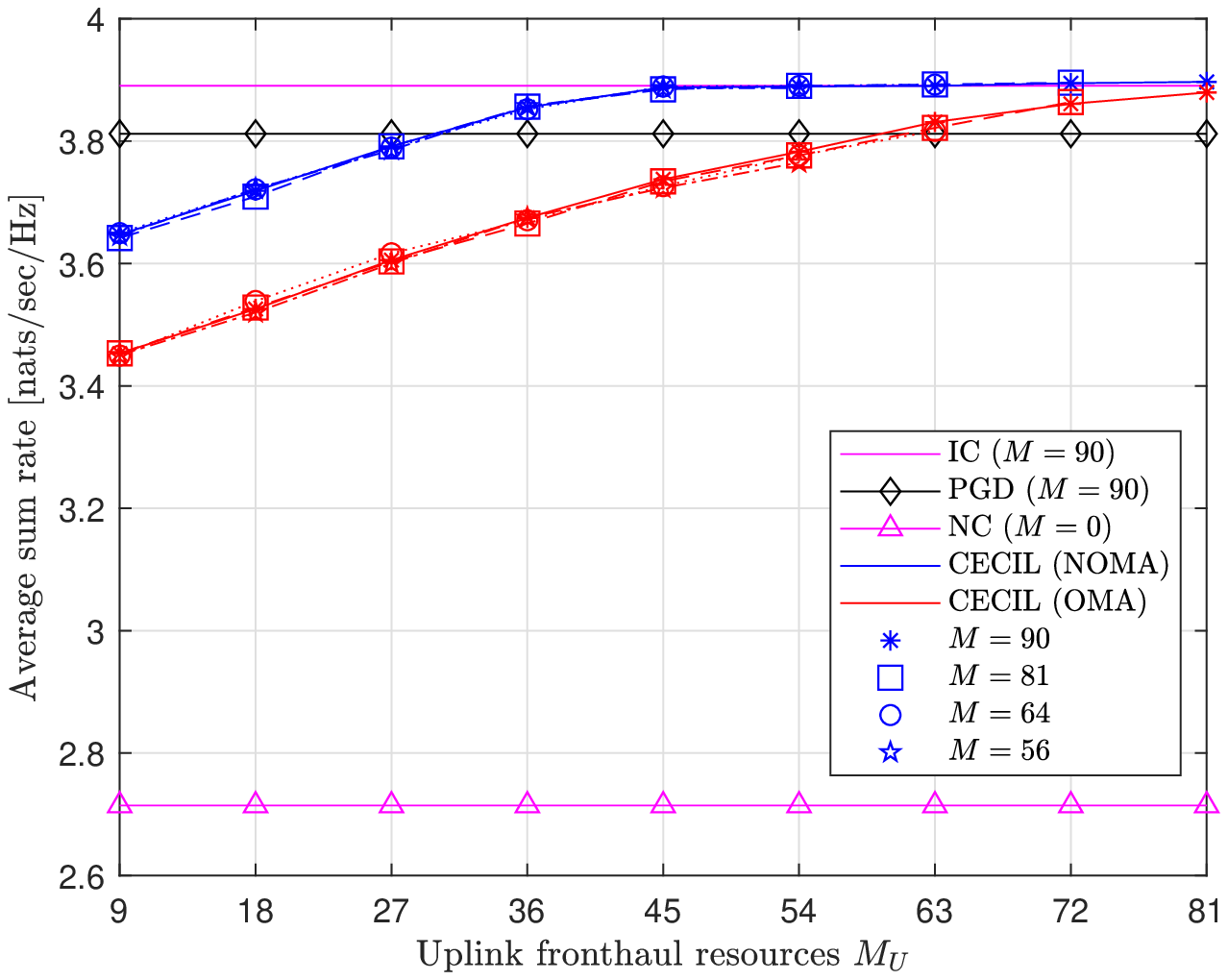}\label{fig:fig4c}
    }
    \caption{Average sum rate performance with respect to $M_{U}$.}
    \label{fig:fig4}
\vspace{-3mm}
\end{figure*}

Fig. \ref{fig:fig4} exhibits the average sum rate performance by changing the number of the uplink fronthaul RBs $M_{U}$ for various choice of the total number of the RBs $M$. For fair comparison with the IC and PGD methods, the maximum of $M$ in the simulations is set to $N(N+1)$. For fixed $M_{U}$ and $M$, the number of the downlink fronthaul RBs is determined as $M_{D}=M-M_{U}$. The OMA system evenly allocates the uplink and downlink RBs for each EN, i.e., $M_{i0}=\frac{M_{U}}{N}$ and $M_{0i}=\frac{M_{D}}{N}$. Fig. \ref{fig:fig4a} depicts the performance with $N=5$ ENs. We can see that the proposed CECIL outperforms the NC benchmark for all simulated $M_{U}$ and $M$, even at a small number of the uplink fronthaul RBs, e.g., $M_{U}=5$. The CECIL with the NOMA fronthauling performs better than that with the OMA scheme. With sufficient $M_{U}$, the CECIL is superior to the existing locally optimum PGD method. As $M_{U}$ increases, the proposed schemes reach the upperbound performance of the IC method. For a fixed $M_{U}$, the performance of the proposed schemes does not improve by increasing $M$, or, equivalently, increasing the number of the downlink RBs $M_{D}=M-M_{U}$. This means that the uplink coordination, which uploads the encoding of the local CSI $\mathbf{a}_{i}$ from the ENs to the cloud, is more crucial than the downlink interaction that forwards the results of the cloud computing to the ENs. Thus, for fixed $M$, the optimum fronthaul resource allocation policy is to assign $M_{D}$ as small as possible, e.g., $M_{D}=N$, and utilize the remaining ones for the uplink coordination as $M_{U}=M-M_{D}$. For the NOMA, $M=20$ RBs with the allocation scheme $M_{U}=15$ and $M_{D}=5$ are sufficient to achieve the performance of the IC requiring $M=30$ RBs, thereby saving $10$ RBs. As expected in Section \ref{sec:sec5C}, a more RBs are needed for the OMA as $M=30$, which is the same as the IC baseline. 

Similar observations are made from Figs. \ref{fig:fig4b} and \ref{fig:fig4c} presenting the sum rate with $N=7$ and $9$ ENs, respectively. We can numerically find that $M=\frac{1}{2}N(N+3N)$ RBs with the allocation $M_{U}=\frac{1}{2}N(N+1)$ and $M_{D}=N$ suffices for the NOMA method to get close to the upperbound IC performance. This is much smaller than $M_{U}=N^2+1$ obtained from the analysis in Section \ref{sec:sec5C}. Compared to the IC and PGD methods requiring $M=N(N+1)$ RBs, the proposed CECIL with the NOMA fronthauling can save total $\frac{1}{2}N(N-1)$ RBs while achieving the same sum rate performance. Still, the OMA method needs $M=N(N+1)$ RBs with $M_{U}=N^{2}$ and $M_{D}=N$. We can conclude that the NOMA fronthauling is more efficient than the OMA for any given $N$ both in terms of the performance and the fronthaul signaling overheads.

\begin{figure*}
\centering
    \subfigure[$N=5$]{
        \includegraphics[width=.33\linewidth]{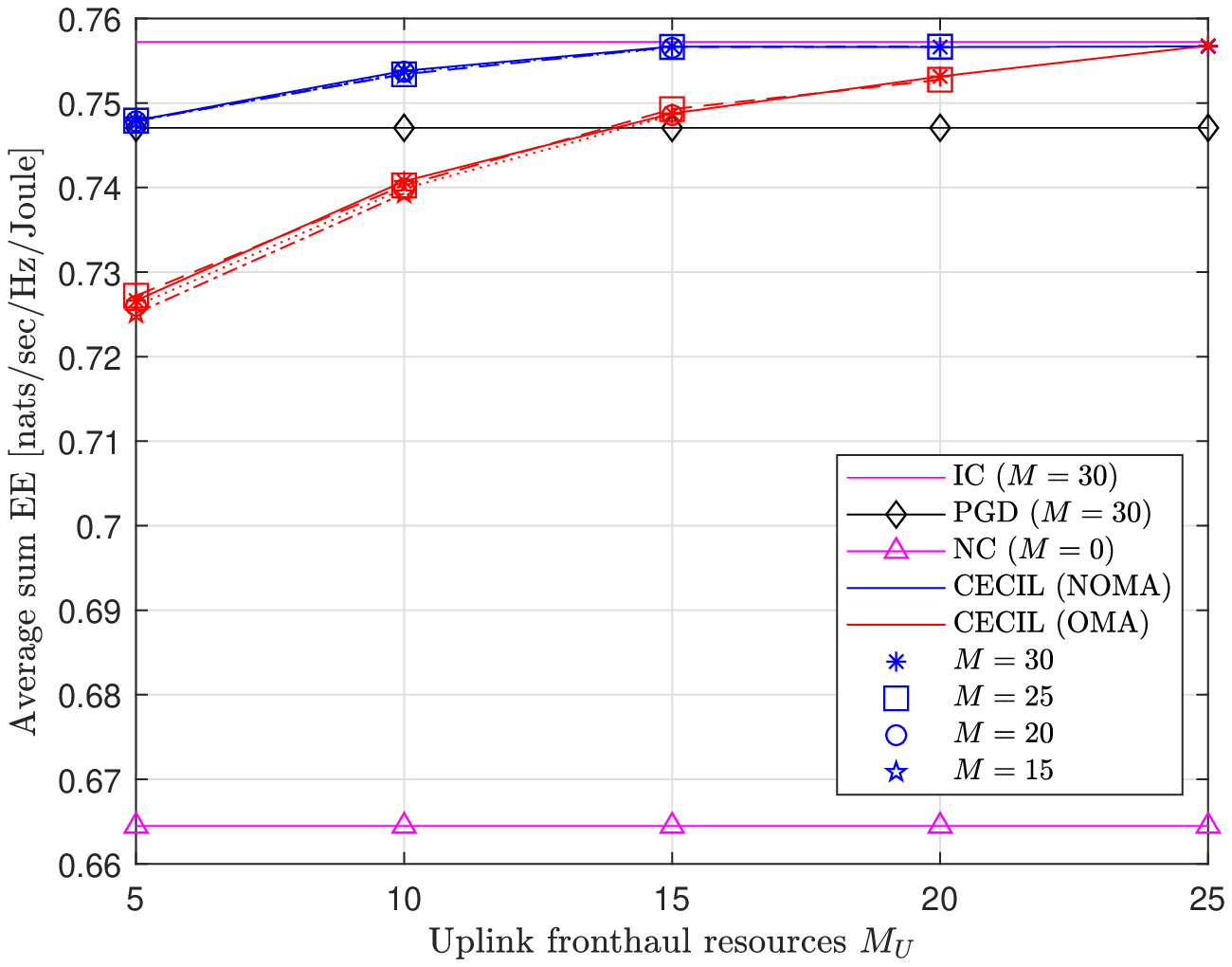}\label{fig:fig5a}
    }\hspace{-5mm}
    \subfigure[$N=7$]{
        \includegraphics[width=.33\linewidth]{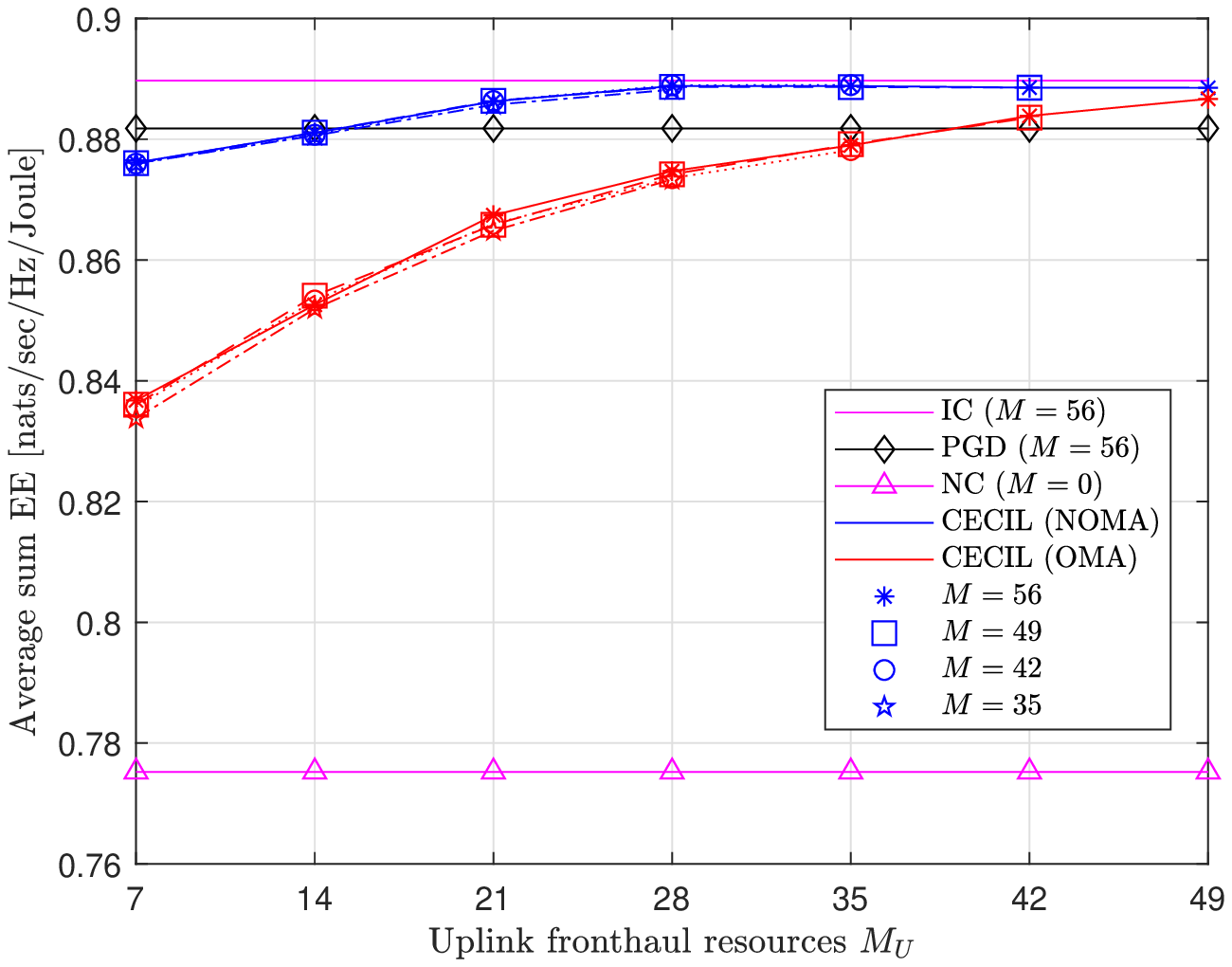}\label{fig:fig5b}
    }\hspace{-5mm}
    \subfigure[$N=9$]{
        \includegraphics[width=.33\linewidth]{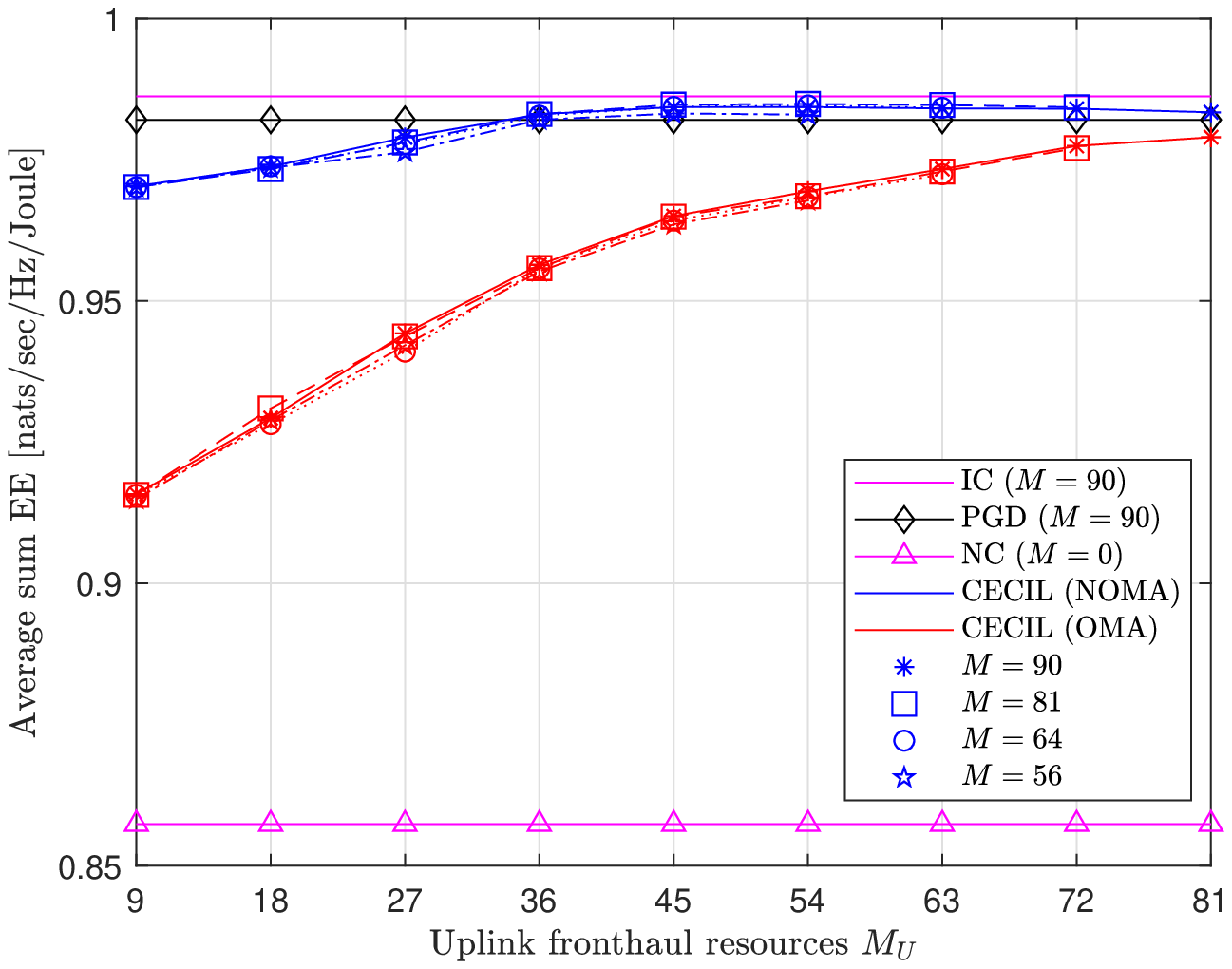}\label{fig:fig5c}
    }
    \caption{Average sum EE performance with respect to $M_{U}$.}
    \label{fig:fig5}
\vspace{-3mm}
\end{figure*}

The EEMax problem is examined in Fig. \ref{fig:fig5} which presents the average sum EE with respect to $M_{U}$. Similar phenomenons to the SRMax results are observed. The proposed approaches work well also in the EEMax formulations and outperforms other baselines. 
It is still beneficial to allocate more RBs to the uplink fronthaul interactions. The NOMA system with $M=\frac{1}{2}N(N+3N)$ RBs for $M_{U}=\frac{1}{2}N(N+1)$ and $M_{D}=N$ achieves the performance identical to the IC method. We can conclude that the CECIL generally performs well for arbitrary utility functions.

%

\begin{table}[]
\caption{Average GPU running time [sec]}\label{tab:tab1}
\vspace{-2mm}
\centering
\begin{tabular}{c|c|c|c|c|c|c|}
\cline{2-7}
                                      & \multicolumn{2}{c|}{$N = 5$} & \multicolumn{2}{c|}{$N = 7$} & \multicolumn{2}{c|}{$N = 9$} \\ \cline{2-7}
                                      & SRMax       & EEMax      & SRMax       & EEMax      & SRMax       & EEMax      \\ \hline
\multicolumn{1}{|c|}{PGD}             & 6.142          & 1.210       & 10.812          & 1.641       & 14.134          & 2.348       \\ \hline
\multicolumn{1}{|c|}{CECIL (NOMA)} & \multicolumn{2}{c|}{0.541}   & \multicolumn{2}{c|}{0.858}   & \multicolumn{2}{c|}{1.310}        \\ \hline
\multicolumn{1}{|c|}{CECIL (OMA)}  & \multicolumn{2}{c|}{0.539}   & \multicolumn{2}{c|}{0.859}   & \multicolumn{2}{c|}{1.308}   \\ \hline
\end{tabular}
\vspace{-3mm}
\end{table}

Table \ref{tab:tab1} compares the online time complexity in terms of the GPU running time for parallel executions of $10^4$ test samples. Both the proposed and PGD methods are implemented with the identical Tensorflow environment to exploit GPU-enabled parallel computations. Both the NOMA and OMA systems employ $M=N(N+1)$ RBs with $M_{U}=N^{2}$ and $M_{D}=N$ which is the same setting to the PGD method. The proposed approaches significantly reduce the GPU running time of the traditional PGD algorithm that requires iterative calculations in the real-time inference. The execution time of the PGD varies for the formulations since its convergence speed highly relies on the structure of the utility functions. The SRMax generally needs a higher computational complexity than the EEMax. On the contrary, the proposed schemes show the identical time complexity performance regardless of the formulations since their online computations depend only on the structure of the DNNs. The result implies that the CECIL framework outperforms the traditional optimization algorithm in terms of the performance, signaling overhead, and computational complexity.

\subsection{Imperfect Fronthaul Link Case}
The rest of this section demonstrates the proposed CECIL method in the imperfect fronthaul link case. For simplicity, we focus on the SRMax with $N=5$ ENs. The noisy fronthaul channels in Section \ref{sec:sec6A} is considered first. The noise vectors in \eqref{eq:yOMA_AWGN} and \eqref{eq:yNOMA_AWGN} are generated as the Gaussian random vectors with zero mean and covariance $\sigma^{2}\mathbf{I}$. The peak power constraint is imposed for the message transmission on each RB. The elements of the message vectors are designed to lie in the bounded range $[-1,+1]$ by applying the hyperbolic tangent activation $\text{tanh}(x)\triangleq\frac{e^{x}-e^{-x}}{e^{x}+e^{-x}}$ to the output layers of the DNNs in \eqref{eq:mi0_fnn} and \eqref{eq:m0i_fnn}. Then, the fronthaul signal-to-noise ratio (SNR) can be defined as $\text{SNR}=\frac{1}{\sigma^{2}}$. 

\begin{figure}
\centering
    \subfigure[NOMA]{
        \includegraphics[width=.4\linewidth]{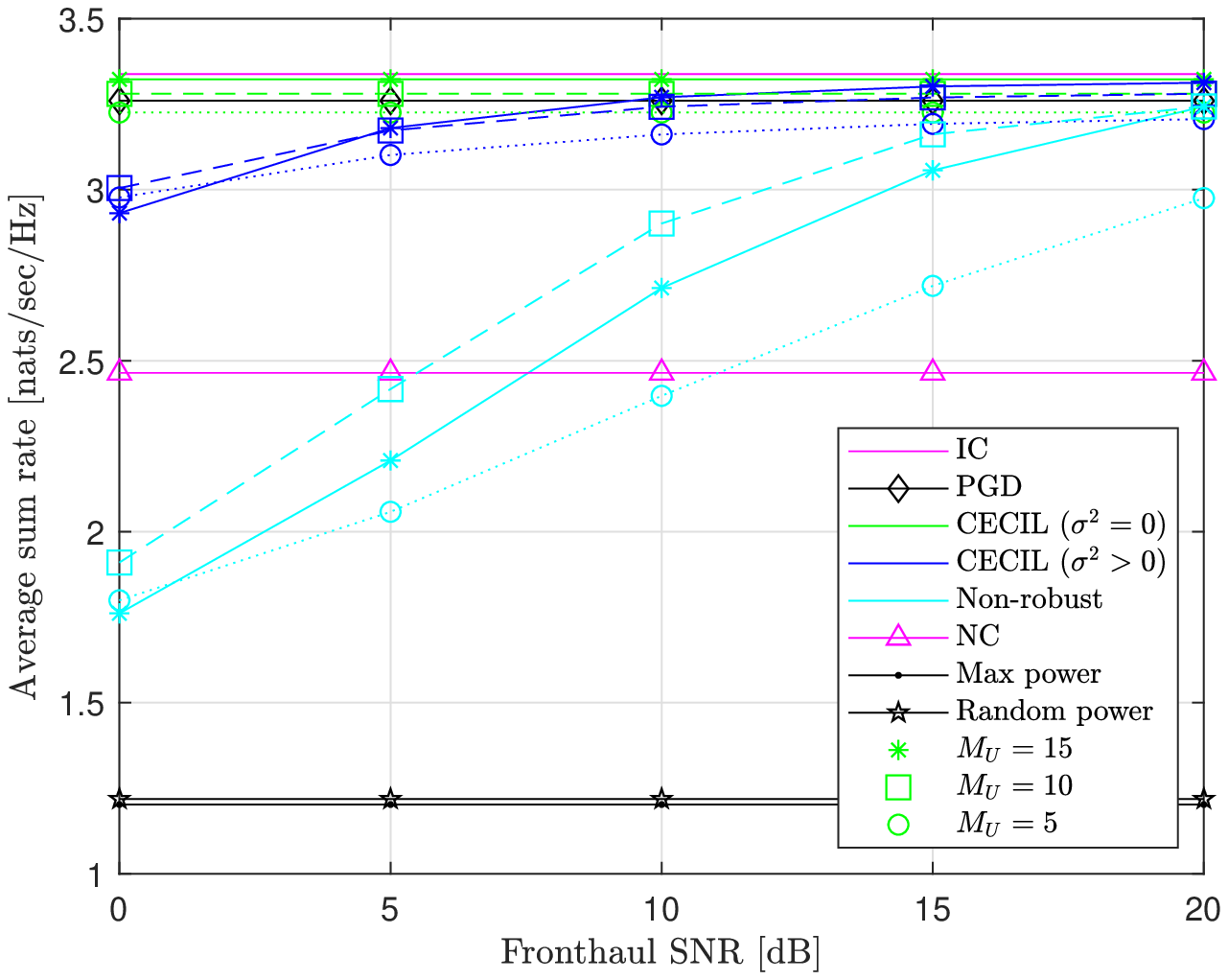}\label{fig:fig6a}
    }
    \subfigure[OMA]{
        \includegraphics[width=.4\linewidth]{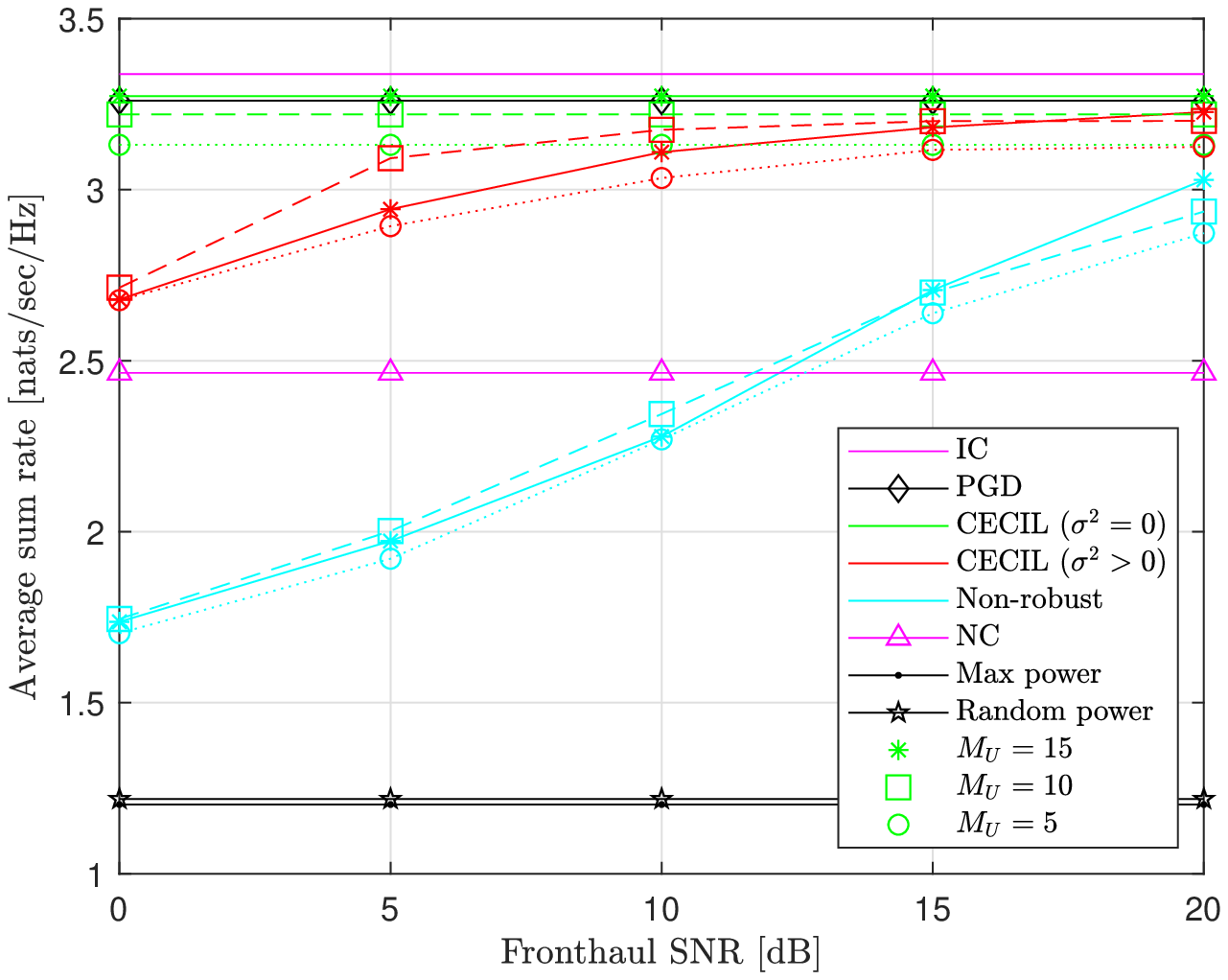}\label{fig:fig6b}
    }
    \caption{Average sum rate performance with respect to SNR for $M=20$.}
    \vspace{-3mm}
    \label{fig:fig7}
\end{figure}

Fig. \ref{fig:fig7} illustrates the average sum rate performance with $M=20$ RBs by changing the fronthaul SNR. For comparison, the performance of the CECIL trained and tested without the noise, i.e., $\sigma^{2}=0$, is plotted. The non-robust scheme stands for the case where the CECIL is trained with the perfect fronthaul links as $\sigma^{2}=0$ but its test performance is evaluated in the presence of the noise. Two naive power control policies, i.e., the max power scheme with $x_{i}=P$ and the random power method with uniformly generated power $x_{i}\in[0,P]$, are also depicted. Both in the NOMA (Fig. \ref{fig:fig6a}) and OMA (Fig. \ref{fig:fig6b}) scenarios, the proposed method converges to the performance of the perfect cooperation case of $\sigma^{2}=0$ as the SNR grows. For all simulated $M_{U}$, the robust CECIL trained with the random noise presents a remarkable performance gain over the NC baseline even in the low SNR regime. This implies that the proposed cloud-aided coordination policy is beneficial for the practical noisy fronthaul channels. The non-robust design exhibits a fairly degraded performance. In the low SNR regime, the performance of the non-robust design becomes worse than the NC method, meaning that the fronthaul cooperation is not helpful if the DNNs are not carefully trained. This verifies the importance of the proposed robust learning strategy which includes random fronthaul noises in the training data set.

\begin{figure}
\centering
\includegraphics[width=.45\linewidth]{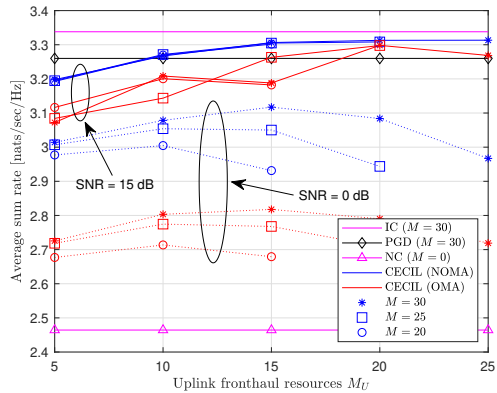}
\vspace{-3mm}
\caption{Average sum rate performance with respect to $M_{U}$.}
\vspace{-3mm}
\label{fig:fig8}
\end{figure}

Fig. \ref{fig:fig8} provides the sum rate performance as a function of $M_{U}$ for the fronthaul SNRs of $5$ and $10\ \text{dBs}$. The NOMA system is still superior to the OMA method in the presence of the noise. In the high SNR regime ($\text{SNR}=15\ \text{dB}$), it is efficient to allocate more RBs to the uplink fronthaul link as in the perfect fronthaul case. This is however not true at $\text{SNR}=0\ \text{dB}$. For fixed $M$, the increase in $M_{U}$ would lead to the performance degradation. There would be a nontrivial tradeoff in the uplink-downlink fronthaul RB allocation for the imperfect fronthaul link scenario.

\begin{figure}
\centering
\includegraphics[width=.45\linewidth]{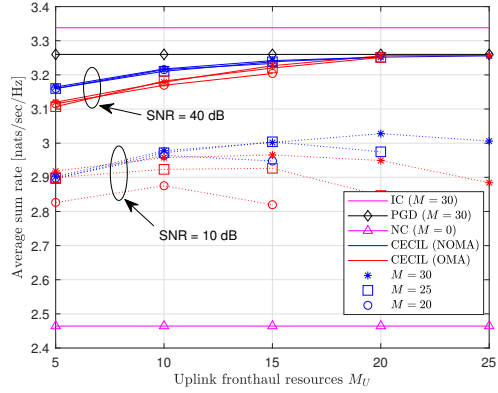}
\vspace{-3mm}
\caption{Average sum rate performance with respect to $M_{U}$ with asymmetric fronthaul channel gains.}
\vspace{-3mm}
\label{fig:figA}
\end{figure}

In Fig. \ref{fig:figA}, we examine the adaptivity of the CECIL framework in a more realistic setup where the fronthaul interactions undergo asymmetric channel gains. Each elements of the message vectors are multiplied by random channel coefficients drawn from the uniform distribution within [0.1, 1]. The NOMA fronthauling scheme still performs better than the OMA method, demonstrating the effectiveness of the resource sharing nature of the NOMA method \cite{WShin:17}. Both the NOMA and OMA require fairly high fronthaul SNR to achieve the performance of the centralized PGD algorithm. A more sophisticated interaction policy would be needed at the cloud and ENs to capture the impact of the asymmetric fronthaul channel gains.


\begin{figure}
\centering
    \subfigure[NOMA]{
        \includegraphics[width=.4\linewidth]{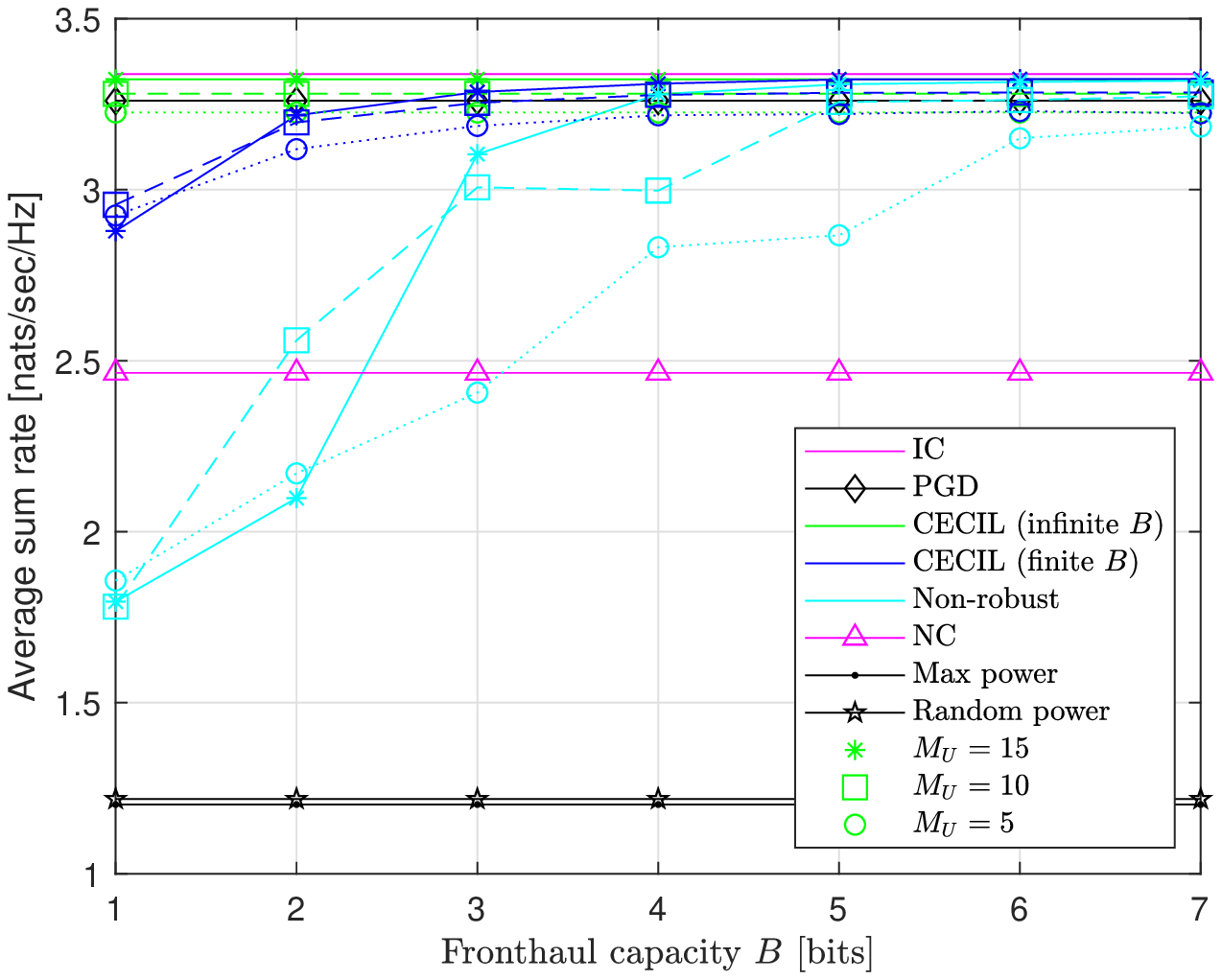}\label{fig:fig9a}
    }
    \subfigure[OMA]{
        \includegraphics[width=.4\linewidth]{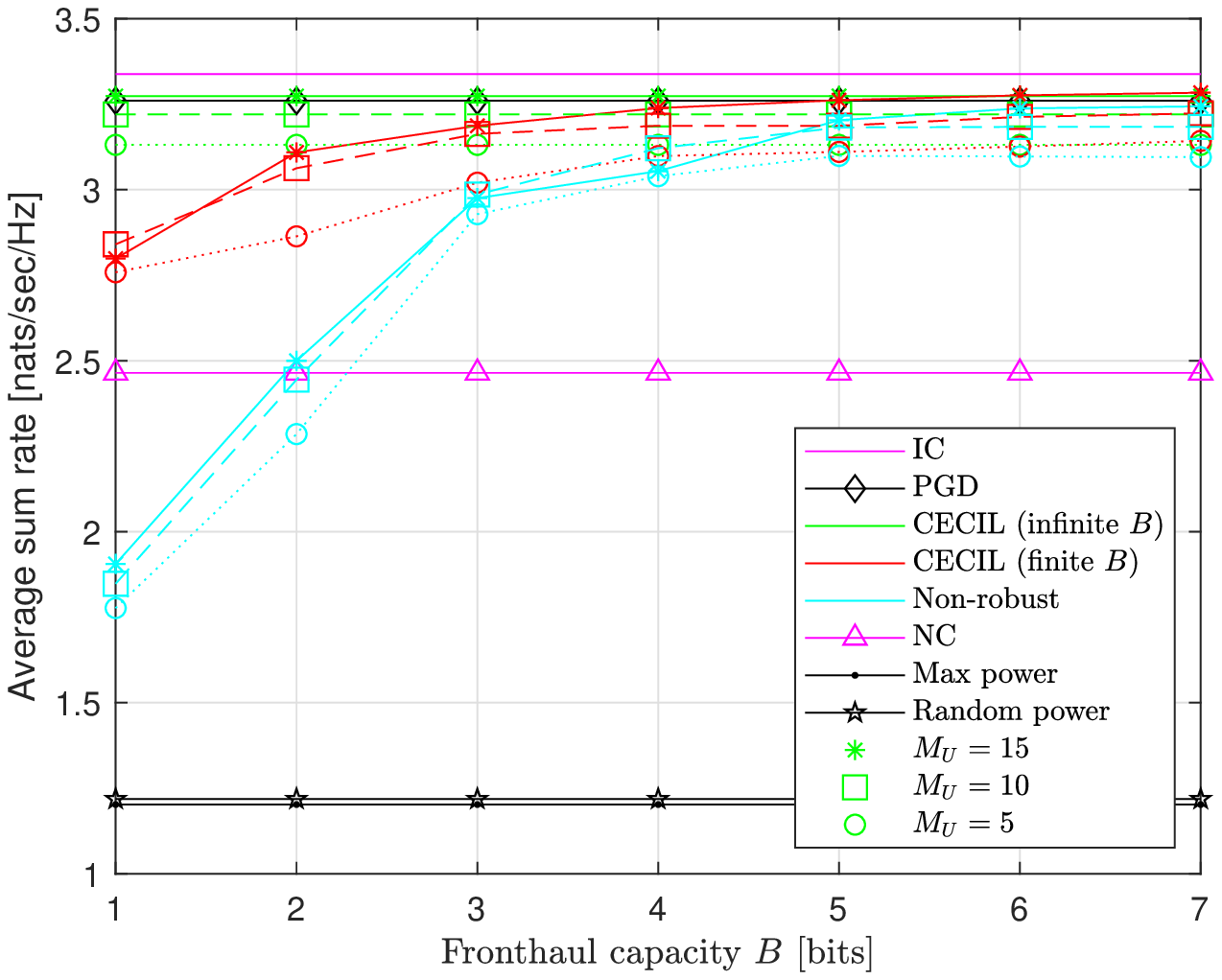}\label{fig:fig9b}
    }
    \caption{Average sum rate performance with respect to $B$ for $M=20$.}
    \vspace{-3mm}
    \label{fig:fig9}
\end{figure}

Next, we investigate the finite-capacity fronthaul link case in Section \ref{sec:sec6B}. The capacity of each fronthaul link is fixed as $C_{il}=2^{B}$, $\forall i,l$, where $B$ reflects the fronthaul capacity in bits.
Fig. \ref{fig:fig9} exhibits the sum rate performance of the finite-capacity fronthaul link case with respect to $B$ for $M=20$ in the NOMA (Fig. \ref{fig:fig9a}) and OMA (Fig. \ref{fig:fig9b}) systems. The performance for the perfect fronthaul link case, i.e., infinite $B$, is shown as a reference. The proposed quantization activation in \eqref{eq:psi} is not included in the non-robust design. Hence, it trains the DNNs in the prefect fronthaul link case, and its test performance is measured with the rounding channel functions with finite $B$. The performance of the proposed message quantization constantly grows as $B$ gets larger and significantly outperforms the non-robust design. We can see that $B=4$ is sufficient for the NOMA system to achieve the upperbound performance of the IC baseline with infinite $B$, whereas the OMA fails to get close to it even with $B=7$ bits.

\begin{figure}
\centering
\includegraphics[width=.45\linewidth]{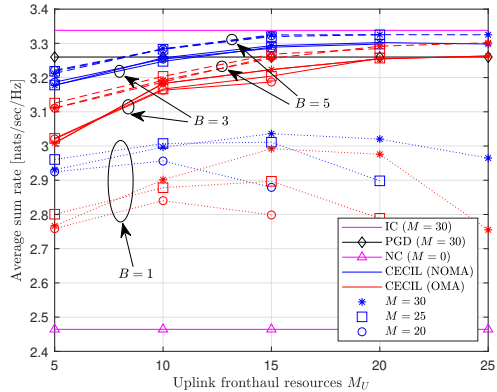}
\vspace{-3mm}
\caption{Average sum rate performance with respect to $M_{U}$.}
\vspace{-3mm}
\label{fig:fig10}
\end{figure}

We plot the average sum rate of the finite-capacity fronthaul case in Fig. \ref{fig:fig10} as a function of $M_{U}$. Similar to the additive noise scenario in Fig. \ref{fig:fig8}, in the finite-capacity case, the optimal fronthaul resource allocation strategy is not trivial if $B$ is small, i.e., the F-RAN suffers from the inaccurate fronthaul interactions. Regardless of $M$ and $B$, the NOMA outperforms the OMA fronthauling scheme in the finite-capacity fronthaul link case. Therefore, we can conclude that the NOMA system is robust to the imperfections incurred in the fronthaul interaction steps.

\section{Concluding Remarks} \label{sec:sec8}
This paper studies a DL solution for addressing generic F-RAN optimization tasks where a cloud schedules decentralized computations of ENs through fronthaul links. A structural learning inference termed by the CECIL framework is proposed which mimics a cloud-aided cooperative optimization strategy. Three different types of DNN modules are applied to the cloud and individual ENs each of which is responsible for uplink and downlink coordinations and distributed optimization. We design message multiple accessing schemes to facilitate the multi-EN fronthaul interactions.
A robust training policy is presented in the practical imperfect fronthaul link scenarios. Numerical simulations validate the superiority of the proposed DL framework over existing optimization algorithms in terms of the performance, fronthaul signaling overheads, and computational complexity. To combat wireless fading fronthaul channels, it would be an interesting future work to adopt channel autoencoder techniques \cite{OShea:17,HLee:19a,HLee:20} for the message-generating inferences. Also, extensions to more complicated application scenarios such as multi-antenna coordinated beamforming problems are worth pursuing.

\bibliographystyle{IEEEtran}
\bibliography{arxiv}


\end{document}